\documentclass[aps,prb,twocolumn,superscriptaddress]{revtex4-2}

\usepackage[dvipsnames]{xcolor}
\usepackage{multirow}
\usepackage{tikz}
\usepackage{graphicx}				
\usepackage{amsmath}
\usepackage{amssymb}
\usepackage{amsfonts}
\usepackage{xspace}
\usepackage{changepage}
\usepackage{mathtools}
\usepackage{verbatim}
\usepackage{hyperref}
\usepackage{bbm}
\usepackage[mathscr]{euscript}

\usepackage[T1]{fontenc} 
\usepackage[utf8]{inputenc}
\usepackage[main=english,french]{babel}

\hypersetup{bookmarksopen=true}

\def\beq{\begin{equation}}
\def\eeq{\end{equation}}
\def\bsp{\begin{split}}
\def\esp{\end{split}}
\def\bea{\begin{eqnarray}}
\def\eea{\end{eqnarray}}
\def\ba{\begin{array}}
\def\ea{\end{array}}
\def\nn{\nonumber \\}

\def\haf#1{{{#1}\over 2}}

\def\l.{\left.}
\def\r.{\right.}

\def\ra{\rangle}

\def\ame{&=&}
\def\ie{{\it i.e. }}

\begin{document}
\title{Frustration, solitons, and entanglement in spin chains}
\author{Christian Boudreault}
\email{christian.boudreault@cmrsj-rmcsj.ca}
\affiliation{D{\'e}partement des sciences de la nature, Coll{\`e}ge militaire royal de Saint-Jean
15 Jacques-Cartier Nord, Saint-Jean-sur-Richelieu, QC, Canada, J3B 8R8}
\affiliation{Groupe de physique des particules, Département de physique,
Université de Montréal,
C.P. 6128, succursale centre-ville, Montréal, 
Québec, Canada, H3C 3J7 }
\author{S. A. Owerre}
\email{alaowerre@gmail.com}
\affiliation{Groupe de physique des particules, Département de physique,
Université de Montréal,
C.P. 6128, succursale centre-ville, Montréal, 
Québec, Canada, H3C 3J7 }
\affiliation{Perimeter Institute for Theoretical Physics, 31 Caroline Street North Waterloo, Ontario, Canada N2L 2Y5 }
\author{M. B. Paranjape} 
\email{paranj@lps.umontreal.ca}
\affiliation{Groupe de physique des particules, Département de physique,
Université de Montréal,
C.P. 6128, succursale centre-ville, Montréal, 
Québec, Canada, H3C 3J7 }
\affiliation{Perimeter Institute for Theoretical Physics, 31 Caroline Street North Waterloo, Ontario, Canada N2L 2Y5 }

\begin{abstract}

Defects in frustrated antiferromagnetic spin chains are universally present in geometrically frustrated systems. We consider the defects of the one-dimensional, spin-$s$ XXZ chain with single-ion anisotropy on a periodic chain with $N$ sites that was famously studied by Haldane. For $N$ odd the antiferromagnetic model is frustrated, and the ground state must include a soliton defect. We consider the Heisenberg interaction perturbatively and determine the corresponding perturbative solitonic ground state. Then we compute the entanglement spectrum, entanglement entropy (EE), capacity of entanglement (CE), and spin correlations in the solitonic ground state. For weak frustration, we find an algebraic violation of the area law for the EE consistent with recent results on weakly frustrated chains. Our analysis then moves beyond the weak frustration regime, and we obtain a novel extensive scaling law for the EE when strong frustration prevails, signalling large entanglement, and failure of the quasiparticle interpretation in this regime. Enhanced frustration results in less total correlations, but relatively more nonlocal correlations.

\end{abstract}

\maketitle

\tableofcontents

\section{Introduction}

The study of phase transitions and entanglement properties of frustrated systems is a venerable subject~\cite{elliot1961phenomelogical, Toulouse1977spin, Vannimenus1977theory, fisher1980infinitely, binder1986spin} which has attracted renewed interest in recent years~\cite{bramwell2001spin, Giampaolo2011characterizing, wolf2003entanglement, marzolino2013frustration, giampaolo2016interplay, giampaolo2019frustration, maric2020quantum, Maric2021resilience}. Frustration refers to the impossibility for the ground state of a many-body system to locally minimize energy. In classical systems, frustration can only arise because of topological obstructions. For instance, the (classical) antiferromagnetic (AF) Ising chain $|J|\sum_i S^z_i S^z_{i+1}$ minimizes the energy of the local interaction terms $|J| S^z_i S^z_{i+1}$ with the Néel order. On a periodic chain of odd length, however, the Néel order cannot be realized and a defect must be present which causes frustration for one or more local terms. This example of a \textit{geometric} frustration is prototypical~: a theorem by Toulouse and Vannimenus implies that a classical system in any dimension is frustrated if and only if it contains loops of odd length that can be mapped to a (frustrated) AF Ising chain~\cite{Toulouse1977spin, Vannimenus1977theory}. Quantum criteria that reduce to the Toulouse-Vannimenus condition in the classical case have been obtained in~\cite{Giampaolo2011characterizing, marzolino2013frustration}. Odd-numbered periodic AF chains are thus the elementary building blocks of geometric frustration, classical or quantum. Contrary to common belief, these systems are known to display boundary-induced orders and transitions \textit{in the bulk} of large chains~\cite{Campostrini2015quantum, Dong2016ACycle, maric2020quantum, Maric2021resilience}. We stress that these effects of frustration and their corresponding degeneracies can be altered dramatically by adding a single site, thus revealing their nonperturbative nature~\cite{giampaolo2019frustration}. Quantum systems can feature another type of frustration due to the impossibility of realizing certain structures of local entanglement on a global scale. An example is the AF Heisenberg chain $|J|\sum_{i=1} \vec{S}_i\cdot\vec{S}_{i+1}$, where $\vec{S}_i=(S_i^x , S_i^y, S_i^z)$, whose local interaction terms have singlet ground state $(|\uparrow_i\downarrow_{i+1}\rangle - |\downarrow_i\uparrow_{i+1}\rangle)/\sqrt{2}$. No global state can possess such local entanglement on every pair of nearest-neighbours, and the system is frustrated. This \textit{quantum} frustration has of course no classical equivalent. It will not be discussed further here.

In this work, we are concerned with the geometric frustration of a quantum spin chain, especially when frustration becomes strong. To define strong frustration, we use the measure of local frustration proposed in~\cite{Giampaolo2011characterizing, marzolino2013frustration}. For a many-body Hamiltonian $H=\sum_S h_S$, frustration of the interactions $h_S$ on subsystem $S$ is a number $f_S\in [0,1]$ quantifying how much a ground state of $H$ is failing to overlap with the ground state of $h_S$. (The quantity $f_S$ will be defined explicitly in Section~\ref{S:EE_res}.) Absence of frustration corresponds to $f_S=0$, while maximal frustration gives $f_S=1$. We define \textit{weak} ($f_S\sim 0$) and \textit{strong} ($f_S\sim 1$) frustration accordingly. Geometric frustration can also be \textit{extensive}. For instance, it is extensive in the AF Ising model on the triangular lattice because the number of frustrated rings (triangles) scales like system size. The systems that we consider in this work have non-extensive frustration, but most have strong frustration. (That is to say, we distinguish between \textit{weak} and \textit{non-extensive} frustration, contrary to~\cite{giampaolo2019frustration}.) Our main results relate to aspects of ground state bipartite entanglement in geometrically frustrated chains, mostly when frustration is strong. Let us briefly recall what bipartite entanglement is, how it can be measured, and why we measure it. For a system with bipartition $\{A, B\}$, a pure quantum state $|\psi\rangle \in \mathscr{H}_A \otimes \mathscr{H}_B$  is said to be \emph{entangled} if it cannot be factored as a product $|\psi_A\rangle \otimes |\psi_B\rangle$ with pure states $|\psi_A\rangle \in \mathscr{H}_A$ and  $|\psi_B\rangle \in \mathscr{H}_B$. A simple example is that of a pair of qubits in the (maximally entangled) state 
\beq
|\psi\rangle=(|00\rangle +e^{i\theta}|11\rangle)/\sqrt{2}.
\eeq
Here, the individual state of qubit $A$ is not itself a pure state, but a mixed state, as can be seen by tracing out qubit $B$, leading to the reduced density operator on $A$,
\beq
\rho_A = \text{Tr}_B |\psi\rangle\langle\psi | = \begin{pmatrix} \frac{1}{2} & 0\\ 0 & \frac{1}{2}\end{pmatrix}.
\eeq
The distribution of eigenvalues of $\rho_A$ reflects the degree of mixity of the latter, which is in turn a good measure of the degree of bipartite entanglement in $|\psi\rangle$. Indeed, a widely used entanglement monotone \footnote{An entanglement monotone is a nonnegative function of a multipartite state which does not increase under the set of local operations and classical communications (LOCC).} for pure states $\rho$ is the bipartite entropy of entanglement (EE), defined as
\beq
S_A = -\text{Tr}_A \:\rho_A \log\rho_A .
\eeq
By inspection, one recognizes that the bipartite EE coincides with the von Neumann entropy of the eigenvalues of $\rho_A$, indicating the degree of mixity of $\rho_A$ (\ie entanglement of $\rho$), with value zero in case $\rho_A$ is pure. For quantum many-body systems, there is now massive evidence that entanglement features in the ground state of these systems, as measured by bipartite EE in particular, are highly sensitive to fundamental properties of the low-lying spectrum like the presence or absence of a mass gap, degeneracies, Fermi surfaces, and criticality~\cite{zeng2015quantum}, properties which may be extremely difficult to assess directly. Moreover, entanglement is a resource for quantum protocols of calculation, communication, and teleportation~\cite{Chitambar2019quantum}. Devising ways to produce highly entangled states in the lab has immediate practical value for these technologies. Although generic quantum states display extensive scaling in the bipartite EE,
\beq\label{E:EE_vol_law}
S_A\sim O(1)\cdot |A|,
\eeq
where $|\cdot |$ stands for set cardinality, systems of local interactions usually present much less entanglement in their ground state. In local lattice systems without frustration, a remarkably common relationship exists between the presence of a spectral gap and the so-called \emph{area law} for the bipartite EE :
\beq
S_A \leq O(1)\cdot |\partial A|,
\eeq 
where $\partial A$ is the frontier of $A$. Exact results are established for spin chains and lattices~\cite{Hastings2007area,Cho2014sufficient}, harmonic lattices~\cite{Plenio2005entropy}, topological phases on lattices~\cite{Michalakis2013stability}, and even for systems with moderate nonlocality~\cite{Kuwahara2020area}. Several applications rely on the area law, e.g. the density matrix renormalization group (DMRG) and matrix product states (MPS), owing to the fact that the area law considerably constrains the complexity of states and systems~\cite{Eisert_2010}. Local spin chains that can be effectively described by a $(1+1)$d conformal field theory ($\text{CFT}_2$) when poised at criticality display a mild, logarithmic violation of the area law~\cite{Holzhey1994geometric}. Frustration may change that picture dramatically. In spin chains with \emph{weak} frustration, one may find (i) an algebraic violation of the area law in the bulk, (ii) a saturation of the bipartite EE as system size is sent to infinity, and (iii) agreement with the non-frustrated case for distances comparable to the correlation length~\cite{giampaolo2019frustration}. Combinations of analytical and numerical approaches indicate that the perturbative picture, in which the frustrated ground state is analyzed in terms of single-particle excitations (defects) over the non-frustrated ground state, sometimes persists beyond the perturbative regime~\cite{giampaolo2019frustration}.

The main objective of the present work is to probe the \emph{strong} frustration regime. We will study ground state entanglement in a quantum spin chain (defined in the next section) with adjustable geometric frustration. Our calculations will be performed by higher-order perturbation theory. The weak frustration regime will be seen to agree perfectly with properties (i)--(iii), and with the single-particle picture. The phenomenology of the strong frustration regime, on the other hand, is radically different, and is our main result. 

The paper is organized as follows. In Section~\ref{S:model}, we present the model and the region of parameter space on which we will concentrate. In Section~\ref{S:results}, we describe our results : the profile of the soliton defect arising from frustration in a classical limit of our model (Section~\ref{S:free_theory_res}), the perturbative solitonic ground state of the interacting model (Section~\ref{S:perturbation_res}), and features of ground state entanglement distinguishing weak and strong frustration (Section~\ref{S:EE_res}). The results are further discussed in Section~\ref{S:discussion}, and we sum up in the Conclusion, Section~\ref{S:conclusion}. Detailed calculations for all our results are provided in the Appendices. 

\section{The model}\label{S:model}

Haldane \cite{haldane} considered the Heisenberg model with an anisotropy, corresponding to the Hamiltonian
\beq\label{E:H}
H=|J|\left(\sum_{i=1}^N \vec{S}_i\cdot\vec{S}_{i+1} +\lambda S^z_iS^z_{i+1}+\mu (S^z_i)^2\right)
\eeq
where $\vec{S}_i=(S_i^x , S_i^y, S_i^z)$, with periodic boundary conditions, $\vec{S}_{N+1}=\vec{S}_1$, for large spin $|\vec{S}|=s\gg 1$ but for small anisotropy $0<(\lambda-\mu)^{1/2}\ll1$, with $\lambda>\mu$. In his work, the low-energy effective field theory was found to be the $O(3)$ nonlinear sigma model, and soliton solutions were semiclassically quantized, showing distinctly different behaviors for integer versus half-integer spin. However, the complete phase diagram of the model, for all values of the couplings, is still of much interest.  We will consider the model in the large anisotropy limit, the opposite limit to that considered by Haldane.  We define $a=|J|\mu$ and $b=|J|\lambda$, and consider the model perturbatively for $|J|\to 0$ but $a$ and $b$ finite.   Thus we write the Hamiltonian as
\bea\label{E:H_ab}
H(\vec{S}_1,\dots ,\vec{S}_N)\ame H_0(S_1^z,\dots ,S_N^z)+|J|\sum_{i=1}^N \vec{S}_i\cdot\vec{S}_{i+1}, \nn
H_0(S_1^z,\dots ,S_N^z)\ame\sum_{i=1}^{N} \Big(a(S_{i}^z)^{2} + bS^z_{i}S^z_{i+1}\Big)\label{h}
\eea
and treat $|J|$ perturbatively. The anisotropy, however small in Haldane's work, picks the antiferromagnetic Néel-ordered ground state that is aligned in the $z$ direction.  For different parts of the parameter space in the anisotropy, it is possible and indeed true that a different ground state is indicated.  The anisotropic term in the Hamiltonian only involves the $z$ component of the spin thus, considering it alone, it is essentially an Ising \cite{Ising} model.  In fact it corresponds exactly to the model studied by Blume and Capel (for spin 1) \cite{bc,CAPEL1,CAPEL2,CAPEL3}, albeit here, it is for arbitrary and large spin.     Therefore we will call the limiting model defined by $H_0(S_1^z,\dots ,S_N^z)$ the Blume-Capel-Haldane-Ising (BCHI) model.

Being a sum of mutually commuting operators, the BCHI Hamiltonian $H_0(S_1^z,\dots ,S_N^z)$  is fully classical. The eigenstates of $H_0(S_1^z,\dots ,S_N^z)$ are obvious and independent of the parameters, $a$ and $b$, and can be labelled as $|s_1,\dots ,s_N\rangle$, where $s_i$ is the $z$ component of the $i$th spin and as usual $s_i\in\{-s,-s+1,\dots ,s-1,s\}$. The corresponding energy eigenvalue is $E(s_1,\dots ,s_N)=\sum_{i=1}^{N} \Big(a(s_{i})^{2} + bs_{i}s_{i+1}\Big)$.  Which eigenstate has the minimum energy, \ie which state is the ground state, is not always obvious. Of special interest to us is the case of frustrated antiferromagnetic coupling on the periodic chain of \emph{odd} length $N$, host to a solitonic defect in the Néel state, as described in the next section.

\section{Results}\label{S:results}

\begin{figure*}
\includegraphics[width=\textwidth]{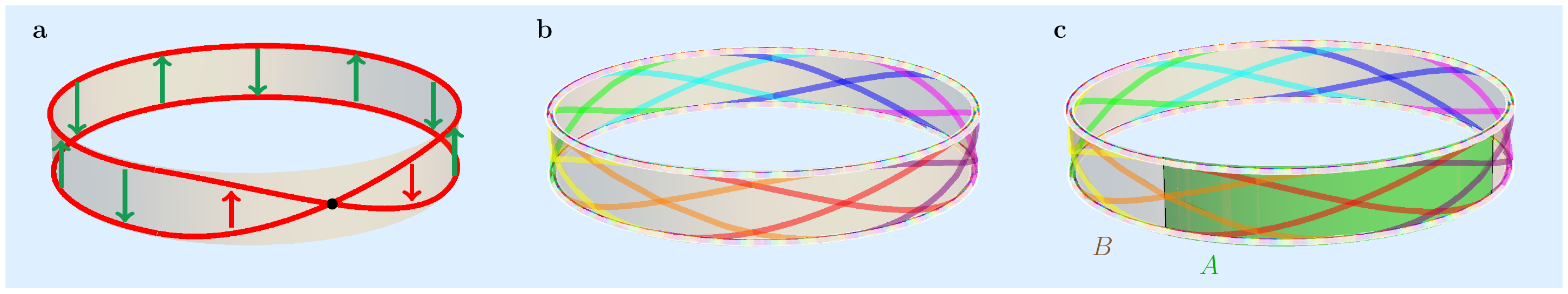}
\caption{(Color online) The main steps of our argument. \textbf{a.} (Section~\ref{S:free_theory_res}) The (semi)classical soliton of the BCHI theory $H_0$ due to frustrated antiferromagnetic coupling on a chain of odd length $N$. (Here, $N=11$, $a/b\sim 0.77$. See also Fig.\ref{F:Spin20}\textbf{c}.) Non-maximal spins are in red and black, see Eqn.\eqref{E:sj_AF_res}, while the Néel background is in green. \textbf{b.} (Section~\ref{S:perturbation_res}) The perturbative solitonic ground state of the full Hamiltonian $H_0+|J|\sum_k\vec{S}_k \cdot\vec{S}_{k+1}$, with small $|J|$. The ground state is a superposition of solitons at all positions, restoring translational symmetry. (The only exception being the special case with half-odd spin $s$, and length-one solitons.) \textbf{c.} (Section~\ref{S:EE_res}) A bipartition of the chain with non-overlapping intervals $A$ and $B$. We study entanglement between these two regions in the perturbative ground state.}\label{F:Fig_summary}
\end{figure*}

We now give an overview of our results as follows. In Section~\ref{S:free_theory_res} we will summarize the main features of the soliton defect present in the (frustrated) BCHI theory with AF couplings and odd number of sites $N$. In Section~\ref{S:perturbation_res} we turn the Heisenberg interaction on, and perform higher-order perturbation to determine the perturbative ground state when the soliton must be present due to frustration. In Section~\ref{S:EE_res} we study entanglement in the perturbative solitonic ground state, and find our main results. These three main steps are illustrated in Fig.~\ref{F:Fig_summary}. The detailed calculations for the results of Sections~\ref{S:free_theory_res}, \ref{S:perturbation_res}, and \ref{S:EE_res} are provided in Appendices~\ref{A:BCHI}, \ref{D}, and \ref{A:EE}, respectively. 

\subsection{Solitons of the classical theory}\label{S:free_theory_res}
We describe the profile of the BCHI soliton for antiferromagnetic coupling ($b>0$), with  odd $N$. To the best of our knowledge, these simple results have not been previously obtained in the literature. Details and calculations may be found in Appendix~\ref{A:BCHI}. Recall the BCHI Hamiltonian,
\beq
H_0(S_1^z,\dots ,S_N^z)=\sum_{i=1}^{N} \Big(a(S_{i}^z)^{2} + bS^z_{i}S^z_{i+1}\Big),
\eeq
with obvious eigenstates $|s_1,\dots ,s_N\rangle$. For easy-axis coupling, $a<0$, the defect is a kink in the Néel background, that is to say, a pair of adjacent parallel maximal spins. However, for easy-plane coupling, $a>0$,  we find that the defect spreads out to maximal size as the antiferromagnetic coupling $b$ is weakened. When the spin is not restricted to quantized values, quadratic optimization gives the semiclassical expression of the soliton extended over sites $j=1,\dots , M$ :
\beq\label{E:sj_AF_res}
s_j =\frac{(-1)^j s}{\sin (M+1)\theta / 2} \;\sin \big(\tfrac{M+1}{2} - j\big)\theta,
\eeq
where $\cos\theta = \frac{a}{b}$, and $M\geq 1$ is the unique integer such that $\cos\frac{\pi}{M+1}<\frac{a}{b}<\cos\frac{\pi}{M+2}$. The rest of the chain is covered by a Néel configuration. The soliton corresponds to a rotation of the (staggered) spin components by $\pi$ radians over the sites labelled by $j=1,\dots , M$, interpolating smoothly between the two ends of the Néel arrangement. The size of the soliton is independent of the number of sites $N$ but depends on the ratio $a/b$, increasing with it as described above.  Hence, the size is a characteristic that is not an artefact of the odd number of sites, and the same soliton can be excited in chains with other lengths and boundary conditions. 

\begin{figure*}
\includegraphics[width=\textwidth]{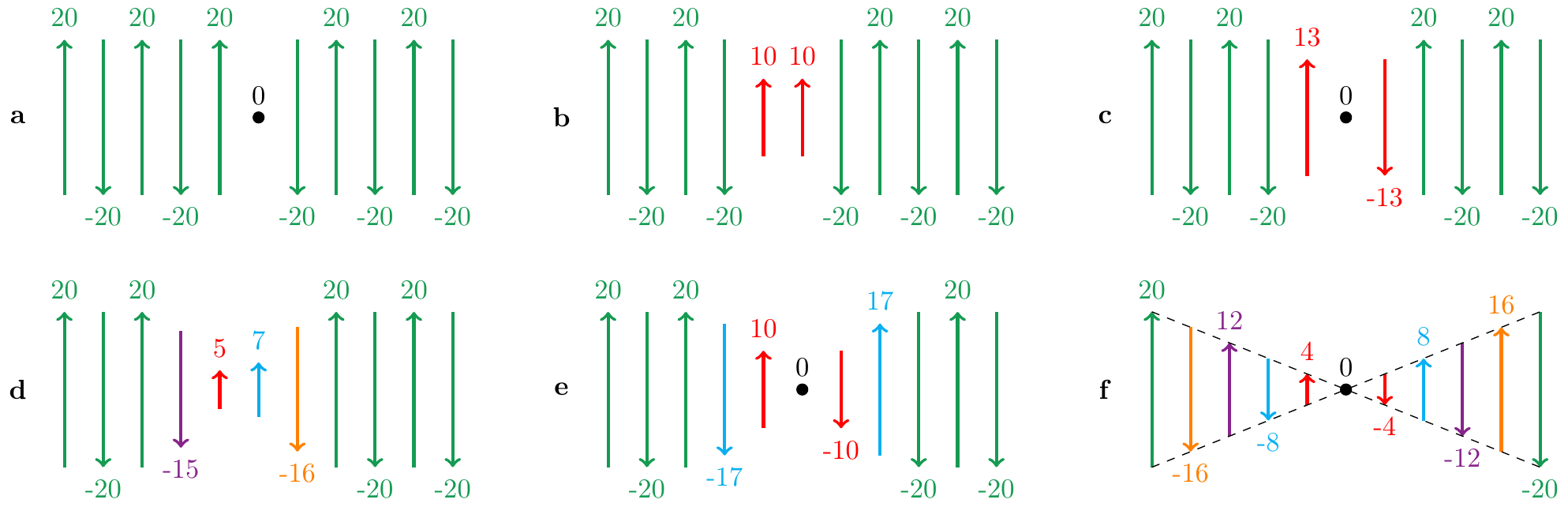}
\caption{(Color online) The soliton of spin 20 on a chain of length $N=11$ for different chosen values of $a/b$. The values of $s^z$ are given in units of $\hbar$. (Numerically evaluated with the \textsc{Mathematica} function \texttt{NMinimize}, with up to 350 iterations using the methods \texttt{DifferentialEvolution} and \texttt{SimulatedAnnealing}.) \textbf{a.} $a/b=0.498$. \textbf{b.} $a/b=0.503$. \textbf{c.} $a/b=0.769$. \textbf{d.} $a/b=0.833$. (Degeneracies are not shown.) \textbf{e.} $a/b=0.870$. \textbf{f.} $a/b=0.999$. The soliton's length is numerically observed to change from $M$ to $M+1$ at $a/b \sim \cos\pi/(M+2)$, in agreement with the large spin calculation. Solitons of odd length have total spin 0. Solitons of even length have (close to) maximal total spin.}\label{F:Spin20}
\end{figure*}

\begin{figure*}
\includegraphics[width=\textwidth]{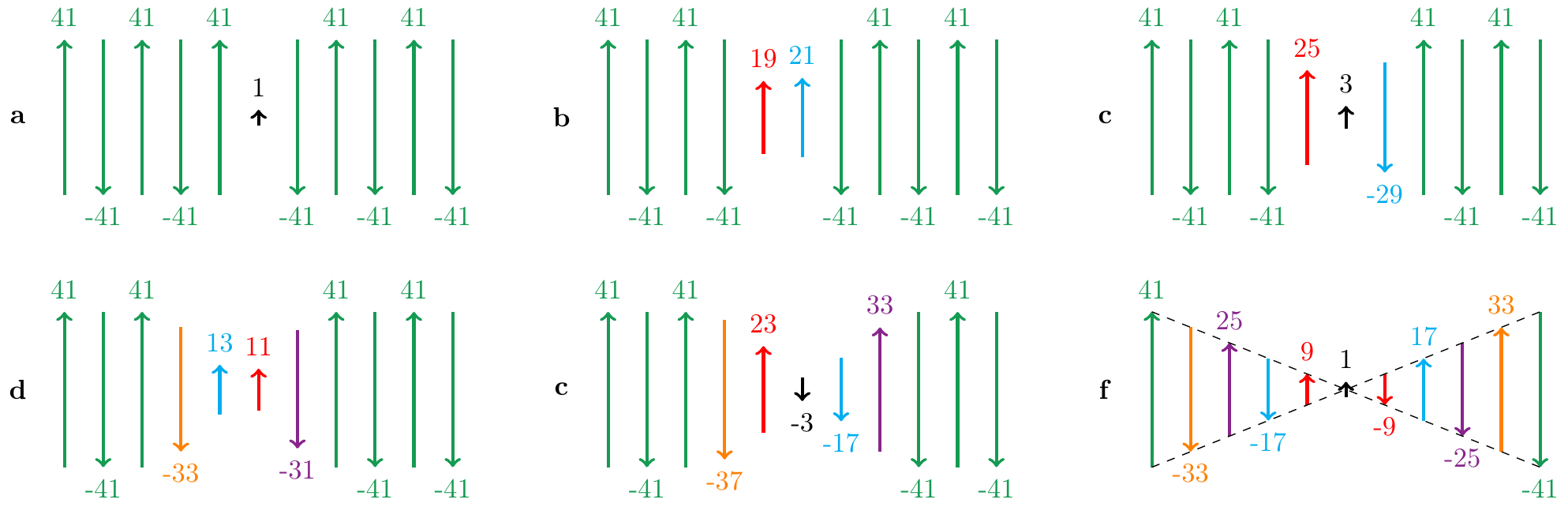}
\caption{(Color online) The soliton of spin $41/2$ on a chain of length $N=11$ for different chosen values of $a/b$. The values of $s^z$ are given in units of $\hbar/2$. (Numerically evaluated with the \textsc{Mathematica} function \texttt{NMinimize}, with up to 350 iterations using the methods \texttt{DifferentialEvolution} and \texttt{SimulatedAnnealing}.) \textbf{a.} $a/b=0.498$. \textbf{b.} $a/b=0.503$. \textbf{c.} $a/b=0.769$. \textbf{d.} $a/b=0.833$. (Degeneracies are not shown.) \textbf{e.} $a/b=0.870$. \textbf{f.} $a/b=0.999$. The soliton's length is numerically observed to change from $M$ to $M+1$ at $a/b \sim \cos\pi/(M+2)$, in agreement with the large spin calculation. Solitons of odd length have total spin 1/2. Solitons of even length have (close to) maximal total spin.}\label{F:Spin41half}
\end{figure*}

When the $z$ component of the spin is quantized, the expression above is only an approximation to the soliton, but is already in good qualitative agreement with numerically obtained solitons of moderate spin. In Fig.\ref{F:Spin20} and Fig.\ref{F:Spin41half}, the soliton is shown with different lengths for spin 20, and spin 41/2, respectively. We observe that the length, the overall symmetry, and the degeneracy of the soliton are quite close to those of the unquantized version. (The quantized antiferromagnetic solitons for spin 1 to 7/2 are provided in the Appendices. See Section~\ref{S:discrete_soliton_appendix}.) Note that when $a/b\lesssim 1$, the soliton has a nearly linear profile. This can be understood analytically from the BCHI Hamiltonian at $a=b$ :
\beq
H_{0,a=b}=\frac{a}{2} \sum_{i=1}^N |S_i^z + S_{i+1}^z|^2,
\eeq
which vanishes on the Néel background, and is minimized when summands $|S_i^z + S_{i+1}^z|$ are all equal on the soliton. 

\subsection{Solitons of the interacting theory}\label{S:perturbation_res} 
From now on, we will focus on the full Hamiltonian, Eqns.\eqref{E:H},\eqref{E:H_ab}, reproduced here
\beq\label{E:H_full}
H=H_0 + |J|\sum_{k=1}^N \vec{S}_k \cdot \vec{S}_{k+1}.
\eeq
As described below, on the chain of odd length $N$ the Heisenberg interaction term will permit soliton translations at high order. Thus, we obtain the solitonic ground state of the frustrated chain perturbatively by allowing soliton translations to restore the translation invariance of the ground state. The procedure is a commonplace application of (higher-order) perturbation theory, in which two elements in the ground space of $H_0$ (two classical solitons at different positions) are connected by the operator $(|J|\sum_k \vec{S}_k \cdot \vec{S}_{k+1})^{\gamma}$ at some order $\gamma$. We have relegated to Appendix~\ref{D} the technical details of the calculations, and give here a brief overview. 

First, we neglect quantum degeneracies not present in the semiclassical picture. For most values of $a/b$, these degeneracies will result from the breaking of a symmetry in the semiclassical expressions, and will be related by the corresponding operators. We thus overlook a number $G_M$ of quantum degeneracies expected to depend of the soliton length $M$, but \textit{not} on the length $N$ of the chain. Specifically, we expect $G_M$ to be upper bounded by the number of quantized configurations within a small distance $\delta$ from the semiclassical soliton in space $[-s,s]^M$, that is $G_M =O(\delta^M)$. One should keep in mind that for any given soliton length $M$, all degeneracies mentioned in the following sections must be multiplied by the corresponding factor $G_M$. 

In order to perform the perturbative level-splitting of the semiclassical soliton ground space, we need to translate the now-interacting solitons by tunnelling. Define the operators
\beq
(S_i^+ S_{i+1}^- )^{\alpha} = \begin{cases}
	(S_i^+ S_{i+1}^- )^{|\alpha|}\quad , \quad \text{if } \alpha\geq 0\\
	(S_i^- S_{i+1}^+ )^{|\alpha|}\quad , \quad \text{if } \alpha < 0.
	\end{cases}
\eeq
Then, for any $s_i, s_{i+1}$, we have
\beq\label{E:adj_trans_res}
(S_i^+ S_{i+1}^- )^{s_{i+1}-s_i}|\dots , s_i , s_{i+1}, \dots\rangle \propto |\dots , s_{i+1} , s_{i}, \dots\rangle. 
\eeq
These adjacent transpositions, including swapping the first and last elements, generate all permutations~\cite{zuylen}, and  all translations in particular. Higher-order Brillouin-Wigner perturbation theory then establishes the ground state for solitons of minimal size (\ie of length one), solitons of intermediate size, and solitons of large size (meaning a size comparable to the total number of sites). 

The ground state corresponds to the superposition of the soliton translated to all positions around the chain, which yields a translationally-invariant ground state (except for half-odd values of $s$ and a soliton of length one, for which translation invariance is broken in accordance with the Lieb-Schultz-Mattis theorem~\cite{Lieb1961two}). Barring that one exception, the ground state has the general form
\beq\label{E:psi_0_res}
|\psi_0\rangle = \frac{1}{\sqrt{N}}\sum_{\mu=1}^{N}\omega^{\mu}|\mu\rangle,
\eeq
where $\omega$ is a root of unity depending on $s$ and $a/b$, and $|\mu\rangle$ is a normal vector parameterized by a classical configuration with soliton at position $\mu$, and Néel background. For solitons of length one and integer spin $s$, for instance, explicit calculation (see Eqn.\eqref{E:gs_evenspin}) gives $\omega=1$ if $s$ is even, and $\omega=-1$ if $s$ is odd. In the above expression, we have omitted an irrelevant overall phase, and we have neglected a small number of $\mathbb{Z}_2$-degeneracies of the (semi)classical soliton, in addition to the already mentioned $G_M$ quantum degeneracies. The effect of those degeneracies on entanglement will be taken care of in Section~\ref{S:EE_res} . For the solitons of length one (\ie one non-maximal spin), the translation by one site is achieved by flipping the unique non-maximal spin with an adjacent spin, which occurs at order $s$ in perturbation.  For intermediate solitons of length $M> 1$, the translation by two sites is achieved at low order, while for a single-site translation, one has to flip the remaining N\'eel state to achieve the translation, a costly operation.  For the large solitons, it is energetically more efficient to translate by a single site, along with the remaining N\'eel part of the chain.

For all ground states of the general form~\eqref{E:psi_0_res}, we find a gapless perturbative spectrum in the form of a band of width $\sim \text{constant}\cdot |J|^{\gamma}$, where the constant depends on the spin $s$ and the parameter ratio $a/b$, and where $\gamma$ is the minimal perturbative order at which soliton translations occur. (See Eqns.\eqref{E:epsilon_length_one},~\eqref{E:epsilonj}, and~\eqref{E:inf_even_epsilon} for the explicit spectra.) This spectrum is very different from what one would obtain without frustration, for then the degenerate classical ground space is spanned by the two Néel states, and the Heisenberg interaction lifts the degeneracy by a finite gap. The spectra for chains of odd length (\ie frustrated) and chains of even length (\ie non-frustrated) are qualitatively distinct, and most importantly, the difference persists in the `thermodynamic limit', if one may still use such terminology in the presence of provably inequivalent limits $\lim_{\text{odd }N\to\infty}$ and $\lim_{\text{even }N\to\infty}$, as resulting from frustration. This phenomenon has also been observed in the geometrically frustrated AF Ising spin chain with a transverse field, which can be solved analytically~\cite{Dong2016ACycle}. For a weak transverse field, these authors also perform (first-order) perturbation theory over the frustrated Ising ground space, which consists of kinks, \ie a single pair of parallel spins in the Néel background, and find a gapless band of translation-invariant combinations of kinks, in perfect agreement with the low-lying states of the analytical result. We note that in our model, kinks are solitons of length $M=0$ (\ie zero non-maximal spin), and correspond to the weakest-frustration scenario, as will be explained in the next section. The novelty of our study is to tackle the \emph{strong} frustration regime. 

Importantly, any energy eigenstate of the perturbative spectrum has the general form~\eqref{E:psi_0_res}. Thus, our entanglement results will not depend on the particular value of $\omega$ corresponding to the ground state. Nevertheless, in Appendix~\ref{D}, we give an attempt at determining the ground state for each soliton length, at the price of using simplifying assumptions when exhaustive calculations are unwieldy. Finally, let us mention that the splitting resulting from the lifting of the classical $\mathbb{Z}_2$-degeneracies and the $G_M$-fold quantum degeneracy does not depend on the (odd) length $N$ of the chain, and will remain finite in the limit $N\to\infty$.

\subsection{Frustration and entanglement}\label{S:EE_res} 
For simplicity, we consider for the rest of this discussion only solitons of length $M\geq 2$ (\ie the region $0<b/2<a< b$ of parameter space). The perturbative ground state of the AF chain of odd length $N$ is then of the general form~\eqref{E:psi_0_res}. We now argue that the frustration of our Hamiltonian, Eqn.\eqref{E:H_ab}, is completely adjustable by tuning the parameter $a/b$. In~\cite{Giampaolo2011characterizing, marzolino2013frustration}, a measure of frustration is proposed for a many-body system $H=\sum_S h_S$ with ground state $|\text{GS}\rangle$, and local interactions $h_S$ on subsystem $S$. Let $\rho=|\text{GS}\rangle\langle\text{GS}|$, and let $\Pi_S\otimes \mathbbm{1}_R$ be the projector onto the ground space of $h_S$, and the identity on the rest of the system $R$. Then $f_S=1-\text{Tr}(\rho\Pi_S \otimes \mathbbm{1}_R)$ quantifies how much $\rho$ fails to overlap with the local subspace selected by $\Pi_S \otimes \mathbbm{1}_R$, and constitutes an unambiguous measure of the frustration of $h_S$~\cite{Giampaolo2011characterizing}. For our Hamiltonian~\eqref{E:H_ab}, the subsystems are neighbour pairs $i,i+1$, and the local interaction terms $h_{i,i+1}$ have ground state $|GS\rangle_i = (|\uparrow\downarrow\rangle \pm |\downarrow\uparrow\rangle)/\sqrt{2}$, up to perturbative corrections~\cite{sp}. Here, the arrows $\uparrow / \downarrow$ denote maximal/minimal $z$-components of spin, $S^z=\pm s$, and the sign in $|GS\rangle_i$ depends on the parity of $2s$. From~\eqref{E:psi_0_res} we find for all sites $i$
\beq
f_{i,i+1}=1-\text{Tr}(|\psi_0\rangle\langle\psi_0|\Pi_{i,i+1}\otimes \mathbbm{1}_R)=\frac{M+1}{N},
\eeq
where the number $M$ of non-maximal spins in the soliton satisfies $2\leq M\leq N-2$. We see that frustration is weak for small solitons, and strong for large solitons, tending to the maximal value 1 when $M/N\to 1$. In the large (odd) $N$ limit, by varying the ratio $a/b$ in the Hamiltonian, frustration covers its entire range of values $(0,1)$. 

\begin{figure}
\includegraphics[width=.48\textwidth]{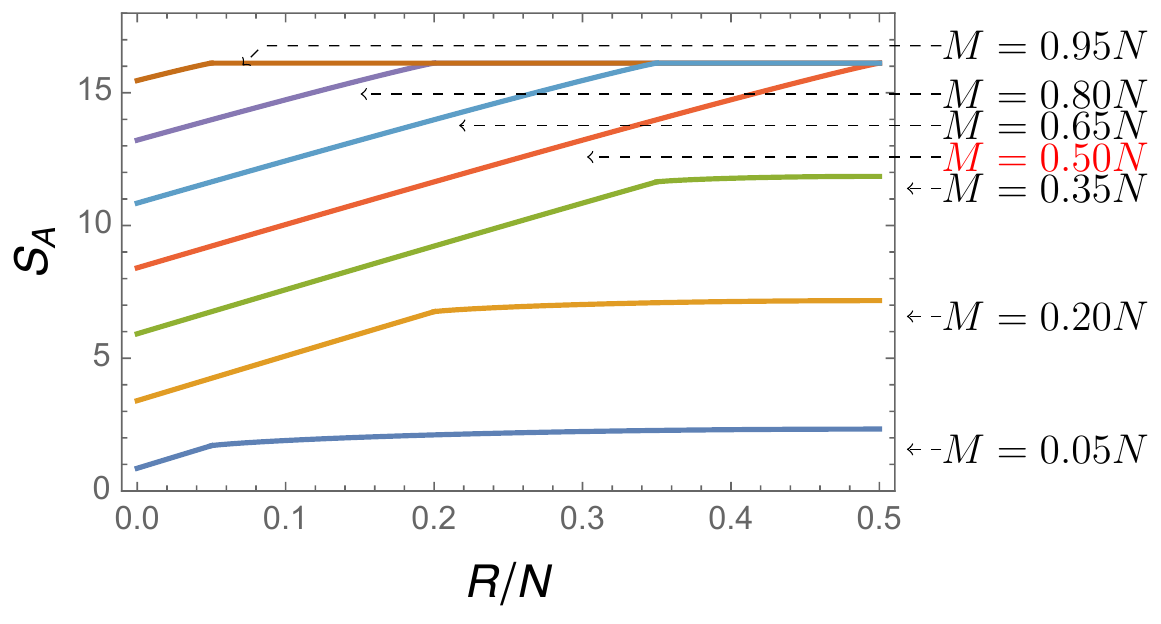}
\caption{(Color online) Entanglement entropy $S_A$ of a connected interval $A$ of size $R$ for the perturbative ground state of the frustrated anisotropic XXZ chain with strong BCHI, and weak Heisenberg. The system size is $N\sim 10^7$, the soliton length ranges from $M=0.05N$ to $M=0.95N$. Observe the different properties corresponding to $M<0.5N$ (small-soliton phase) and $M>0.5N$ (large-soliton phase). }\label{F:EE_soliton}
\end{figure}

\begin{figure}
\includegraphics[width=.4\textwidth]{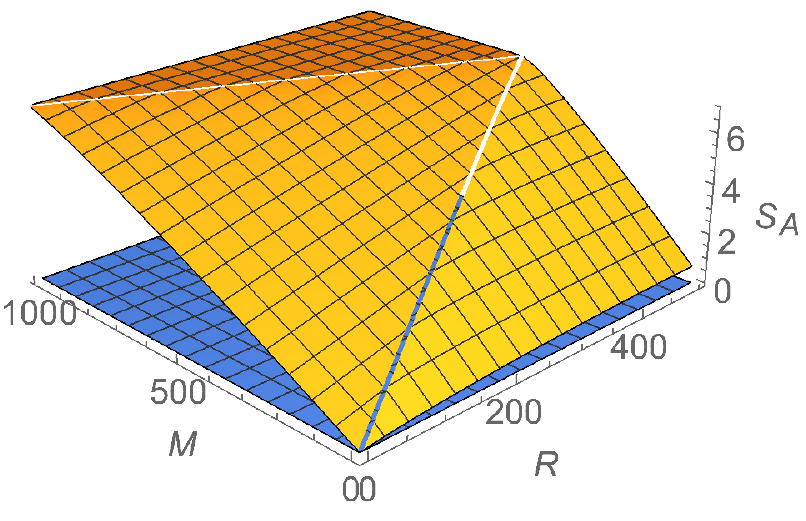}
\caption{Entanglement entropy $S_A$ in the solitonic ground state, for subsystem lengths $R\in[1,N/2]$ and soliton lengths $M\in[2,N-2]$. Here we have used system size $N=1001$. The behaviour is qualitatively different in regions I (right), II (front), and III (top). The EE is extensive in region II.}\label{F:EE}
\end{figure}

In order to study \emph{bipartite} EE, let $A$ and $B$ be two intervals such that $\{A,B\}$ is a bipartition of the chain. In Appendix~\ref{A:EE}, we compute the reduced density operator $\rho_A=\text{Tr}_B|\psi_0\rangle\langle\psi_0|$ corresponding to the ground state~\eqref{E:psi_0_res}, and the resulting bipartite EE, $S_A = -\text{Tr}_A\; \rho_A \ln \rho_A$. We find
\begin{widetext}
\beq\label{E:EE_discrete_res}
S_A=\begin{cases}
\frac{N-M -R +1}{N}\ln \frac{N}{N-M -R +1}+\frac{R-M+1}{N}\ln\frac{N}{R-M +1}+\frac{2M -2}{N}\ln N\quad &,\quad \text{region I}\\
\frac{N-M -R +1}{N}\ln \frac{N}{N-M -R +1}+\frac{M +R -1}{N}\ln N\quad &,\quad \text{region II}\\
\ln N\quad &, \quad \text{region III},
\end{cases}
\eeq
\end{widetext}
with regions I, II, and III as represented in Fig.~\ref{F:regions_res}. (In fact, our calculations yield the full entanglement \emph{spectrum} of the model. See Appendix~\ref{A:EE} for details.)
\begin{figure}[ht]
\includegraphics[width=.26\textwidth]{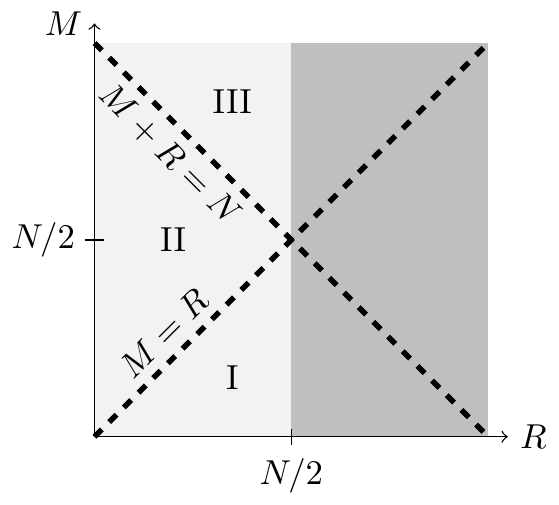}
\caption{The $(R,M)$-space for subsystem length $R$ and soliton length $M$. In the light gray region ($R\leq N/2$) the EE is given by~\eqref{E:EE_discrete_res}. In the dark gray region ($R>N/2$) the EE is obtained by the symmetry $S_{A^c}(R)=S_A (N-R)$, where $A^c$ is the complement of $A$. The entanglement entropy has qualitatively distinct behaviours on regions I, II, and III.}\label{F:regions_res}
\end{figure}
We remind the reader that in the ground state~\eqref{E:psi_0_res} we have neglected a small number of classical $\mathbb{Z}_2$-degeneracies, as well as a number $G_M$ of quantum degeneracies, expected to be upper bounded by the number of quantized configurations within a small distance $\delta$ from the classical soliton in space $[-s,s]^M$, that is $G_M =O(\delta^M)$. As long as transition amplitudes between degenerate states are sufficiently small, the effect of the degeneracies on the entanglement entropy will be a subleading additional term $\log G_M=O(M\log\delta)$, plus a small integer multiple of $\log 2$. 
The entropy is represented in Figs.~\ref{F:EE_soliton} and~\ref{F:EE} for multiple subsystem sizes and soliton sizes. 

\subsubsection{Weak frustration}
Let us first consider weak frustration, $f_{i,i+1}=\frac{M+1}{N} \sim 0$. If the limit of (odd) chain length $N\to \infty$ is reached while $M$ is fixed, we get the \textit{binary entropy} function $S_A = -\frac{R}{N}\ln\frac{R}{N}-\left(1-\frac{R}{N}\right)\ln \left(1-\frac{R}{N}\right)$ corresponding to a randomly positioned particle (soliton) being in $A$ with probability $\frac{R}{N}$, and outside of $A$ with probability $\left(1-\frac{R}{N}\right)$. This weak frustration scenario, $M=O(1)$, is in perfect agreement with the results obtained for the (weakly) frustrated chains of Ref.\cite{giampaolo2019frustration}, namely : (i) a bulk violation of the area law~\cite{Holzhey1994geometric, vidal2003entanglement, Calabrese2004entanglement, Hastings2007area} $S_A=a(N)R^{b(N)}$, where $b(N)\approx 0.22(2)$ (sub-extensive) for $N$ chosen between 201 and 901, and (ii) the saturation of the EE in the limit of large $N$. The universal EE curve that the authors of~\cite{giampaolo2019frustration} identify in the scaling thermodynamic limit of their models is the binary entropy function mentioned above, in agreement with the single-particle interpretation described by these authors as well as in this paragraph. (See also \cite{Castro-Alvaredo2018entanglement, Castro-Alvaredo2018entanglementFF} for a more general quasiparticle interpretation of the EE.) When both $M$ and $R$ are fixed as $N\to \infty$, the dependence on boundary conditions disappears and $S_A\to 0$, in agreement with the non-frustrated chain (open or closed) and its Néel ground state, a product state without entanglement. (If we take care of the $\mathbb{Z}_2$-degeneracy, the ground state is a translation-invariant combination of the two Néel states, and $S_A=\log 2$.)

\subsubsection{Strong frustration}
We now turn to strong frustration, $f_{i,i+1}\not\sim 0$, i.e. $M=O(N)$. If $M$, $R$, and $N$ all grow at the same rate as the limit $N(\text{odd})\to\infty$ is reached, we drop the bounded terms in Eqn.\eqref{E:EE_discrete_res}, and find
\beq\label{E:S_A_continuum_v2}
S_A\sim\left\{\begin{aligned}
\tfrac{2M}{N}\ln N \quad &, \quad \text{region I}\\
\tfrac{R+M}{N}\ln N \quad &, \quad \text{region II}\\
\ln N \quad &, \quad \text{region III}.
\end{aligned}\right.
\eeq
This entanglement behaviour is strikingly different from the one found in the weakly frustrated case, and is our main result. On the one hand, the EE is extensive in subsystem size when $R$ is smaller than the soliton length $M$ (region II), then plateaus as $R$ becomes larger than $M$ (region I). On the other hand, the EE is nonfinite as $N\to\infty$, diverging like $\ln N$ in all regions. This behavior appears irreconcilable with a quasiparticle interpretation as given in \cite{Castro-Alvaredo2018entanglement, Castro-Alvaredo2018entanglementFF}. Interestingly, even though the soliton ground state $|\psi_0\rangle$ has arguably large entanglement when $M\gg 1$ (being extensive in subsystem size), it is also arguably slightly entangled in the sense of algorithm theory. Because the EE diverges no more than $O(\ln N)$, if the solution of a quantum $N$-qubit problem is encoded in $|\psi_0\rangle$, a theorem by Vidal shows that this problem is likely to be efficiently simulatable classically~\cite{Vidal2003efficient}.

\section{Discussion}\label{S:discussion}
We note that the strongly frustrated chain cannot admit a long-distance effective field theory which is scale invariant because the soliton length is a macroscopic physical scale of the system. This is also seen, for weak and strong frustrations, at the level of EE scaling with respect to subsystem length, where we observe algebraic scaling in the frustrated chain, as opposed to logarithmic scaling in $\text{CFT}_2$'s~\cite{Holzhey1994geometric, vidal2003entanglement, Calabrese2004entanglement} and $(1+1)D$ Lifshitz theories~\cite{Ardonne2004topological, Chen2017gapless}. Anticipating the results of the next section, let us mention that the capacity of entanglement can also diagnose the impossibility of a conformal scale-invariant effective limit. However, an aspect of the right effective theory might be found in Fig.~\ref{F:EE_soliton}, where we observe a duality between solitons lengths $M$ and $N-M$ in the plateauing of EE. We identify two perturbative quantum phases, depending on the coupling ratio $a/b$ through the value of the soliton length $M$. The \textit{small-soliton} phase corresponds to $M< N/2$~:
\beq\label{E:phase1}
S_A \sim \left\{\begin{aligned}
\tfrac{R+M}{N}\ln N\quad & ,\quad R< M\quad (\text{extensive})\\
\tfrac{2M}{N}\ln N\quad & ,\quad R> M \quad (\text{plateaued}).
\end{aligned}\right.
\eeq
In perturbation theory, this phase corresponds to soliton translations over two lattice constants, as briefly explained below Eqn.\ref{E:psi_0_res}. (See also Appendix~\ref{S:any_length}.) The \textit{large-soliton} phase corresponds to $M> N/2$~:
\beq\label{E:phase2}
S_A \sim \left\{\begin{aligned}
\tfrac{R+M}{L}\ln N\quad & ,\quad R< N-M\quad (\text{extensive})\\
\ln N\quad & ,\quad R> N-M \quad (\text{plateaued}).
\end{aligned}\right.
\eeq
In perturbation theory, this phase corresponds to soliton (and Néel background) translations over a single lattice constant. (See Appendix~\ref{S:infinite_length}.)  These equations can be brought into a unique form by considering the \textit{relative} entanglement entropy $S_A'(R) = S_A(R) - S_A(0)$, where $S_A(0)=\frac{M}{N}\ln N$ is the EE of an interval of length $O(1)$ in the limit $N,M\to\infty$. (Corresponding to a divergent single-point entropy in the continuous limit.) We find 
\beq\label{E:EE_shifted}
S'_{A,\text{strong frustration}} \sim \left\{\begin{aligned}
\tfrac{R}{N}\ln N\quad & ,\quad R< \lambda\quad (\text{extensive}),\\
\tfrac{\lambda}{N}\ln N\quad & ,\quad R> \lambda \quad (\text{plateaued}),
\end{aligned}\right.
\eeq
where the length scale $\lambda$ is the soliton length $M$ in the small-soliton phase ($M <N/2$), and the soliton colength $N-M$ in the large-soliton phase ($M >N/2$). In the next section, we compare this strong-frustration EE with the behavior of the EE in certain nonlocal field theories.

\subsection{Nonlocal theories}
Some strongly coupled nonlocal field theories~\cite{Karczmarek2013holographic, shiba2014volume, Rabideau2015perturbative} present a particular volume-law scaling of the EE that has recently attracted interest. These theories have in common that they possess a nonlocality scale $\tilde\lambda$ independent of the UV cutoff. In~\cite{shiba2014volume}, for instance, the nonrelativistic scalar field
\beq
H_{\text{nonlocal}}=\int d^D x \;\frac{1}{2}\left((\partial_t \phi)^2 +b\phi\cdot e^{a(-\sum_i\partial_{i}^2)^{w/2}}\cdot\phi\right)
\eeq
has positive constants $a,b,w$ and a nonlocality scale $\tilde\lambda=a^{1/w}$ characterizing the coupling of the field at distant positions. The leading divergence in entanglement entropy has the form
\beq\label{E:nonlocal_vol_law}
S_{A,\text{nonlocal}}\sim\begin{cases}
	|A|\Lambda_{\text{UV}}^{D} & (\text{subsystem scale}\ll\tilde\lambda)\\
	\tilde\lambda |\partial A| \Lambda_{\text{UV}}^{D} & (\text{subsystem scale}\gg\tilde\lambda),
	\end{cases}
\eeq
where $|A|$ is the volume of $A$, $|\partial A|$ is area the boundary, and $\Lambda_{\text{UV}}$ is the momentum scale at the UV cutoff~\cite{Rabideau2015perturbative}. In~\cite{Karczmarek2013holographic}, the `dipolar' nonlocal deformation of super-Yang-Mills $\mathcal{N}=4$ also has a nonlocality scale $\tilde\lambda$ independent of the UV cutoff, and its EE follows~\eqref{E:nonlocal_vol_law}. (The other nonlocal deformation considered by these authors, the `noncommutative' deformation, has a nonlocality scale $\tilde\lambda(\Lambda_{\text{UV}})$ which diverges when $\Lambda_{\text{UV}}\to\infty$, leading to a pure volume law~\eqref{E:EE_vol_law} for the EE.) Intriguingly, the coefficients of the nonlocal volume law~\eqref{E:nonlocal_vol_law} at $D=1$ (in which case $|\partial A|$ is an integer) are reminiscent of the strong-frustration EE that we found in Eqn.\eqref{E:EE_shifted}, but with a crucial difference in their respective divergences. Let us be more specific. For a \emph{physical} frustrated XXZ chain with a fixed, dimensionful lattice constant $\epsilon$, the physical size of interval $A$ is $|A|=R\epsilon$, and the physical scale corresponding to $\lambda$ is 
\beq\label{E:lambda_phys}
\tilde\lambda=\lambda\epsilon.
\eeq
Note that, $\epsilon$ being fixed, the scale $\tilde\lambda$ depends only on the parameter ratio $a/b$. A low-energy effective QFT for the chain would thus have a scale $\tilde\lambda$ independent of the cutoffs, a natural IR cutoff equal to the inverse physical length of the chain, 
\beq\label{E:QFT_IR}
\Lambda_{\text{IR}}=(N\epsilon)^{-1},
\eeq
and a UV cutoff of the order of the inverse lattice constant, $\Lambda_{\text{UV}}\sim\epsilon^{-1}$. For the effective QFT of the chain, the strong-frustration law~\eqref{E:EE_shifted} is thus embodied in the following form :
\beq\label{E:EE_shifted_cont}
S'_{A,\text{strong frustration}} \sim \left\{\begin{aligned}
|A|\Lambda_{\text{IR}}\ln\tfrac{\Lambda_{\text{UV}}}{\Lambda_{\text{IR}}}\quad & ,\quad |A|< \tilde\lambda,\\
\tilde\lambda\,\Lambda_{\text{IR}}\ln\tfrac{\Lambda_{\text{UV}}}{\Lambda_{\text{IR}}}\quad & ,\quad |A|> \tilde\lambda.
\end{aligned}\right.
\eeq
Eqns.\eqref{E:nonlocal_vol_law} and~\eqref{E:EE_shifted_cont} both possess an extensive regime $S_A\propto |A|$, and a plateaued regime $S_A\propto \tilde\lambda$ for some length scale $\tilde\lambda$ of the Hamiltonian.
Both are IR finite (\ie finite as $\Lambda_{\text{IR}}\to 0$). However, Eqn.\eqref{E:EE_shifted_cont} is logarithmically UV divergent, with $S'_{A,\text{strong frustration}} \sim O(\ln\Lambda_{\text{UV}})$ as $\Lambda_{\text{UV}}\to \infty$, in stark contrast with the more severe UV divergence $O(\Lambda_{\text{UV}})$ of the nonlocal volume law~\eqref{E:nonlocal_vol_law} for $D=1$. A precision is in order concerning the ultraviolet limit of the effective QFT of the chain. This QFT, taken by itself, has a cutoff-independent scale $\tilde\lambda$ in its Hamiltonian, and happens to be the low-energy limit of the class of frustrated XXZ chains whose values of $N,\epsilon$ and $a/b$ are in agreement with the IR cutoff of the QFT, as per Eqn.\eqref{E:QFT_IR}, as well as with the scale $\tilde\lambda$ of the QFT, as given by Eqn.\eqref{E:lambda_phys}. The UV limit of the effective QFT, with fixed $\Lambda_{\text{IR}}$ and $\tilde\lambda$ values, consists in the sequence of chains with increasingly small lattice constant $\epsilon$, and values of $N$ and $a/b$ corresponding to the fixed values $\Lambda_{\text{IR}}$, $\tilde\lambda$. 

As already observed, the leading UV divergence of the effective QFT's EE at strong frustration is only logarithmic, in contrast to the nonlocal volume law~\eqref{E:nonlocal_vol_law}. Nevertheless, we believe that the similarity between~\eqref{E:nonlocal_vol_law} and~\eqref{E:EE_shifted_cont} is remarkable. Could it be an indication that, despite the difference, perhaps the low-energy effective field theory over the solitonic ground state displays some form of nonlocality, with the soliton length (or colength) as the nonlocality scale? This question, however, is beyond the scope of the present work. 

\subsection{Capacity of entanglement}\label{S:CE}
\begin{figure}
\includegraphics[width=.4\textwidth]{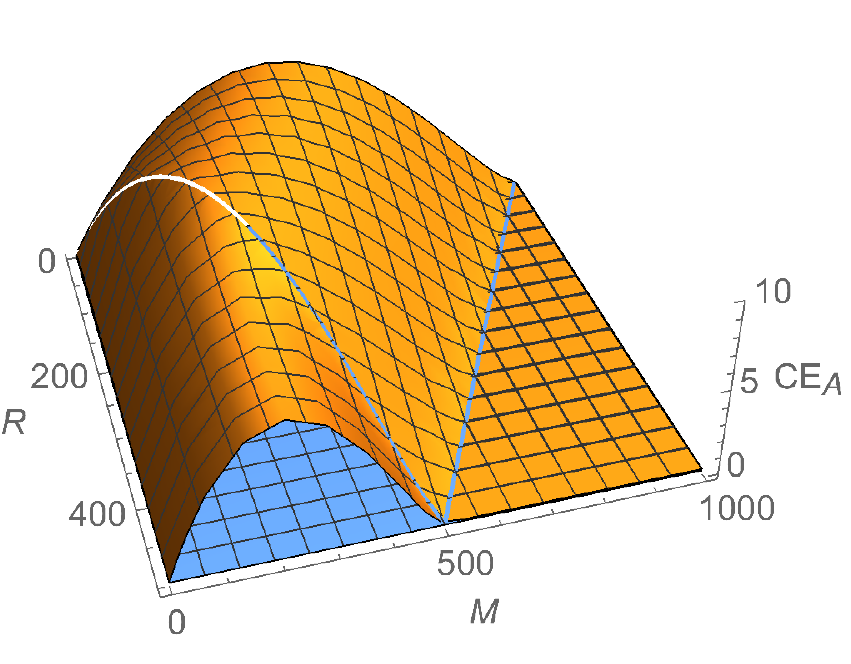}
\caption{The capacity of entanglement $CE_A$ in the solitonic ground state, for subsystem lengths $R\in[1,N/2]$ and soliton lengths $M\in[2,N-2]$. The system size is $N=1001$. The behaviour is qualitatively different in regions I (left), II (top), and III (right). The CE vanishes in region III.}\label{F:CE}
\end{figure}

\begin{figure}
\includegraphics[width=.45\textwidth]{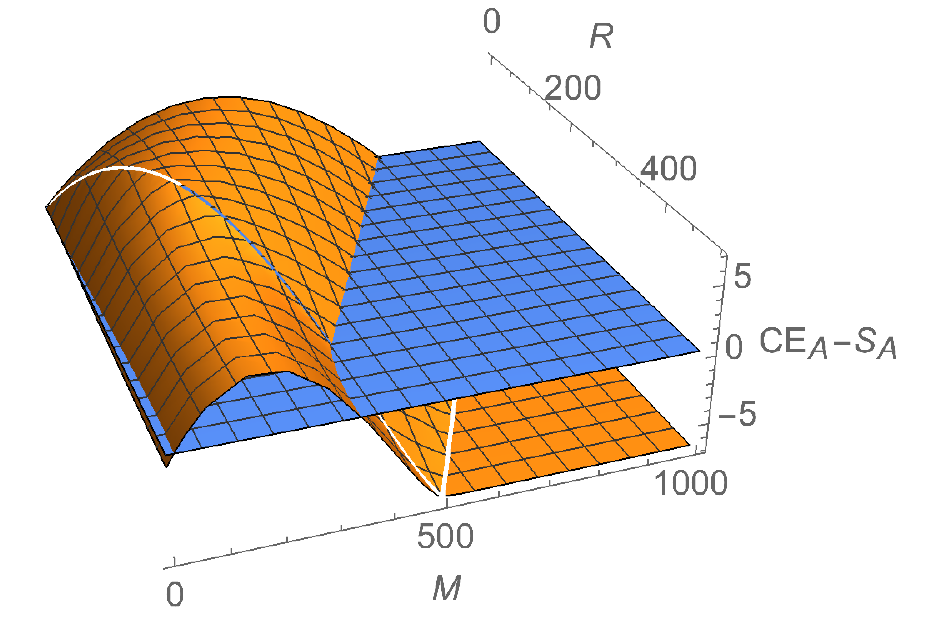}
\caption{The difference $CE_A - S_A$ in the solitonic ground state, for subsystem lengths $R\in[1,N/2]$ and soliton lengths $M\in[2,N-2]$. The system size is $N=1001$. We see that $CE_A \sim S_A$ (i.e. $CE_A- S_A \sim 0$ for all $R$ and fixed $M$) only when $M\sim O(1)$, confirming that the frustrated system has no conformal low-energy limit.}\label{F:CEminusEE}
\end{figure}

The capacity of entanglement (CE) is another quantity associated to a reduced density matrix, defined in the same way as one defines heat capacity for thermal systems~\cite{Yao2010entanglement, deBoer2019aspects}. (The reduced density matrix $\rho_A$ corresponding to the perturbative solitonic ground state is calculated in Appendix~\ref{A:EE}.) From the modular Hamiltonian $H_A$ defined as
\beq\label{E:modular_H_res}
\rho_A = \sum_n e^{-\xi_n}|n\rangle_A {}_A\langle n | =e^{-H_A}\quad ,
\eeq
with eigenvalues $\xi_n$, one defines the capacity of entanglement of subsystem $A$ as the variance of $H_A$,
\beq
CE_A = \left(\sum_n \xi_n^2 \; e^{-\xi_n}\right) - S_A^2= \langle H_A^2\rangle - \langle H_A \rangle^2=\text{var}(H_A).
\eeq
The CE is therefore a measure of the width of the eigenvalue distribution for the modular Hamiltonian, and for the reduced density matrix. From the explicit modular Hamiltonian of the solitonic ground state (see Eqn.\eqref{E:H_A}), we may compute de CE of any single-interval $A$. We find
\begin{widetext}
\beq
CE_A\sim\left\{\begin{aligned}
\tfrac{N-M-R+1}{N}(\ln \tfrac{N}{N-M-R+1})^2+\tfrac{R-M+1}{N}(\ln \tfrac{N}{R-M+1})^2+\tfrac{2M-2}{N}(\ln N)^2 - S_A^2 \quad &, \quad \text{region I}\\
\tfrac{N-M-R+1}{N}(\ln \tfrac{N}{N-M-R+1})^2+\tfrac{R+M-1}{N}(\ln N)^2 - S_A^2 \quad &, \quad \text{region II}\\
0 \quad &, \quad \text{region III},
\end{aligned}\right.
\eeq 
\end{widetext}
where regions I, II, and III still refer to those defined in Fig.~\ref{F:regions_res}. Note that $CE_A$ is identically zero in region III because all eigenvalues of the reduced density matrix are equal to $1/N$. In this region, the solitonic ground state has the EE and CE of the generalized GHZ state $N^{-1/2}\sum_{i=1}^N |i\rangle^{\otimes N}$. We provide a plot of the CE in Fig.~\ref{F:CE}. Interestingly, the capacity of entanglement in CFTs is found to scale like the entanglement entropy~\cite{deBoer2019aspects}. Moreover, there is convincing evidence that such a scaling, $CE_A\sim S_A$, can detect criticality in many models~\cite{Yao2010entanglement, deBoer2019aspects}. For the solitonic ground state, we observe in Fig.~\ref{F:CEminusEE} that $CE_A \not\sim S_A$ everywhere, except for $M=O(1)$, for which case we have already found a super-logarithmic violation of the area law, not consistent with a low-energy CFT limit. The capacity of entanglement thus confirms that the frustrated chain has no conformal effective QFT.

\subsection{Correlations}
We conclude this discussion with a few words about the effect of frustration on correlations in the solitonic ground state. It is not difficult to determine approximate expressions for the correlator
\beq
C_{N}^{zz}(R)=\langle\psi_0 | S_i^z S_{i+R}^z |\psi_0\rangle = \frac{1}{N}\sum_{i=1}^N \langle\mu |S_i^z S_{i+R}^z |\mu\rangle,
\eeq
where $|\psi_0\rangle$ is as in Eqn.\eqref{E:psi_0_res}. In the last term, obtained from the translation invariance of $|\psi_0\rangle$, $\mu$ is (any) fixed position on the chain. Let us first describe correlations over distances $R > M$. Using the symmetries of the semiclassical soliton, we can show that
\beq\label{E:corr_weak}
C_{N}^{zz}(R)=(-1)^R s^2 \left(1-\frac{2R}{N}\right) \hspace{1cm},\hspace{1cm}R>M.
\eeq
See Appendix~\ref{A:EE} for more details. This expression is valid with any amount of frustration, as long as $R>M$. Notice that antipodal sites, $R\sim N/2$, have vanishingly small correlations. When $M=0$ (weakest frustration), the soliton is a kink, and the solitonic ground state is formally identical with the ground state of the geometrically frustrated AF Ising model, whose correlation function is known exactly~\cite{Dong2016ACycle}, and coincides with~\eqref{E:corr_weak}. As observed in~\cite{Dong2016ACycle}, the function $C_{N}^{zz}(R)$ reveals nonlocal correlations on the frustrated chain, because setting the correlation to a fixed value, and letting $N\to\infty$, will force $R$ to diverge as well. Without surprise, the purely local part of the correlator, $\lim_{N\to\infty}C_{N}^{zz}(R)=(-1)^R s^2$, is that of a simple antiferromagnet. The algebraic decay of correlations is consistent with a high level of entanglement in the ground state.

At moderate or strong frustration, $M\gg 1$, one needs to distinguish the case $R>M$, with correlations given by~\eqref{E:corr_weak}, and the case $R<M$. For the latter case, reasonable assumptions give
\beq\label{E:corr_largeM}
C_{N}^{zz}(R)\approx (-1)^R\; \frac{s^2}{N}\left[N-\frac{2M}{3}-\frac{2R^2}{M}+\frac{2R^3}{3M^2}\right],\; R<M.
\eeq
\begin{figure}
\includegraphics[width=.49\textwidth]{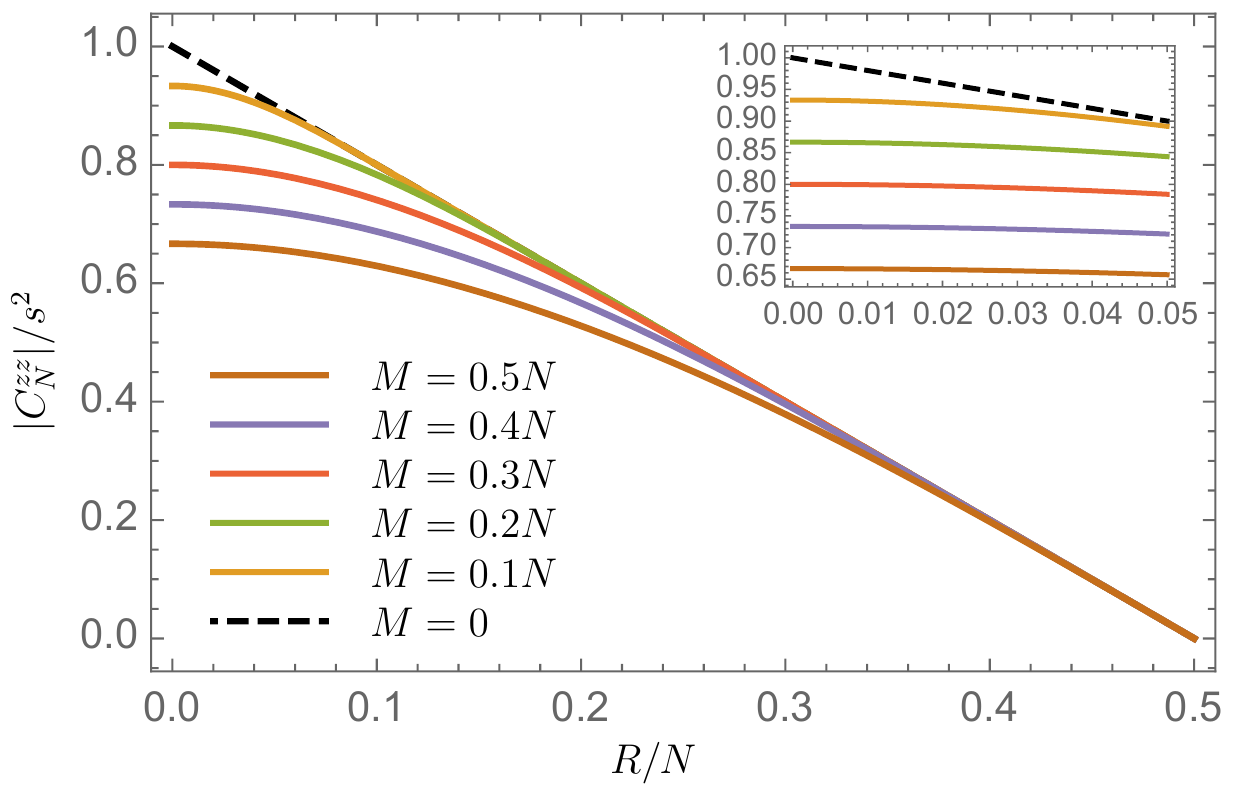}
\caption{(Color online) Dimensionless $zz$ correlations $|C_N^{zz}|/s^2$ as a function of spin separation $R/N$ for different values of soliton length $M$, \ie different values of frustration. The behavior at distances $R<M$ is given by~\eqref{E:corr_largeM}, whereas the behavior at distances $R>M$ is given by~\eqref{E:corr_weak}. Frustration diminishes and flattens correlations within the length of the soliton : at fixed separation $R<M$, the (anti)correlation $|C_N^{zz}|$ and the decay $|\partial C_N^{zz}/\partial R|$ both decrease with $M$. Increased frustration implies relatively more nonlocal correlations. \textbf{Inset.} Correlations between neighboring spins for different values of $M$.}\label{F:corr}
\end{figure}
A plot of $C_{N}^{zz}(R)$ over the full range $R\in[0,N/2]$, given in Fig.\ref{F:corr}, shows that the soliton length $M$ is indeed a characteristic scale of the model. (Anti)correlations are diminished over distances less than the soliton size $M$, and their distribution is flattened. At fixed separation $R<M$, greater frustration (\ie larger soliton size $M$) implies weaker (anti)correlation $|C_N^{zz}|$, and damped decay $|\partial C_N^{zz}/\partial R|$. For close neighbors $R\sim 1$, in particular, we obtain
\beq
C_{N}^{zz}(R)\sim(-1)^R\; s^2\left(1-\frac{2M}{3N}\right),
\eeq
revealing correlations almost independent of separation, but weaker than is found at weak frustration. It thus appears that frustration implies a reducing of the total amount of correlations, affecting primarily the most local correlations, while leaving nonlocal correlations unabated. Simply put, more frustration means \emph{relatively} more nonlocality. A relative dominance of long-distance correlations is suggestive of a highly entangled state, as previously observed in the extensive region of the strong-frustration EE, Eqn.\eqref{E:EE_shifted}, where the nonlocal characteristic scale $\lambda$ was already identified to the soliton length $M$ (or colength $N-M$ in case $M>N/2$, by symmetry). At distances greater than the nonlocal scale, spin correlations are independent of the amount of frustration, and given by~\eqref{E:corr_weak} in all cases. This is in contrast to the EE, which is highly sensitive to frustration at any subsystem size, as seen in Fig.\ref{F:EE}.
We should remark that these results are not exclusive to the solitonic ground state, but apply equally well to the other states of the perturbative spectrum.

Interestingly, frustration acts in seemingly opposite directions on the states of the solitonic perturbative spectrum, reducing (local) correlations while increasing entanglement. As observed above, these trends may be reconciled if we understand them as resulting from enhanced relative nonlocality. Entanglement is a resource for quantum communication, whereas correlations may be a resource for evesdroppers, and be detrimental to the confidentiality of the communication. States with less correlations than their entanglement would suggest are therefore appealing for secure quantum protocols~\cite{Hayden2004randomizing, hastings2007entropy}. It would be of value to determine if, and how, frustration could be harnessed for such purpose.

\section{Conclusion}\label{S:conclusion}
In this paper, we have studied a large-spin Haldane-like anisotropic XXZ model in the limit of large anisotropy which we call the classical, Blume-Capel-Haldane-Ising (BCHI) limit. When geometric frustration is present due to antiferromagnetic coupling between an odd number of sites with periodic boundary conditions, soliton defects are to be found. We have determined the profile of the soliton in the classical limit, and we have computed the perturbative corrections due to the Heisenberg interaction. In this way, we have found the perturbative solitonic ground state of the chain with frustration, along with a continuous band of excited states, in stark contrast with the gapped spectrum prevailing when no frustration is present. Importantly, this phenomenon persists as system size is sent to infinity. That it is sometimes impossible to define a single thermodynamic limit independently of the boundary conditions has been observed in other models as well~\cite{Campostrini2015quantum, Dong2016ACycle,giampaolo2019frustration}. 

Using the measure of frustration introduced in~\cite{Giampaolo2011characterizing, marzolino2013frustration}, we have shown that the amount of frustration increases with the length of the soliton, so that frustration is a tunable parameter in our model, making it possible to overtake the weak frustration regime studied in all previously mentioned references, and probe the effect of \emph{strong} frustration on ground state entanglement. We have determined the entanglement spectrum, entanglement entropy (EE), capacity of entanglement (CE), and spin correlations in the solitonic ground state. For weak frustration, we have found an algebraic violation of the area law for the EE consistent with recent results on weakly frustrated chains~\cite{giampaolo2019frustration}. Moving beyond the weak frustration regime, we were able to reveal that the EE has extensive scaling in subsystem size when frustration is strong. In that regime, we also observe that the EE scales logarithmically with the length $N$ of the chain. We have noticed a similarity (as well as a major difference) between the strong-frustration EE of our model, and the ground state EE of certain recently studied nonlocal field theories~\cite{Karczmarek2013holographic, shiba2014volume, Rabideau2015perturbative}. The nature of the effective low-energy QFT of the frustrated chain is an open problem, but the length of the soliton is bound to be an important scale of it. Remarkably, although entanglement increases with frustration, we have found that more frustration results in a reduction of the total amount of correlations, along with an enhancement of the relative nonlocality of these correlations. 

In a future work, we intend to numerically study the entanglement in randomly generated 3D configurations of spin-$s$ XXZ chains hosting a soliton, using the results obtained in the present work. It will be interesting to determine how the EE of this 3D system scales with subsystem size, soliton length $M$, average length $\langle N\rangle$ of the chains, and chain density. It would also be stimulating to identify frustrated 3D systems hosting the BCHI soliton. These results could shed light on the physics of geometrically frustrated 3D systems, where frustrated AF chains are bound to be found.
Frustration is likely to be very common in nature, and non-extensively frustrated antiferromagnetic 1D chains are universally present in systems with geometric frustration, even in higher dimensions, and whether frustration is extensive or not~\cite{giampaolo2019frustration}. Systems with extensive frustration, including the ANNNI model~\cite{elliot1961phenomelogical, fisher1980infinitely}, spin ices~\cite{bramwell2001spin}, and spin glasses~\cite{binder1986spin}, possess properties strikingly different from those of non-frustrated systems, and the physics of frustrated 1D chains is a promising avenue to better understand this fascinating subject.

\section{Acknowledgments} 
We thank Ian Affleck, William Witczak-Krempa, Clément Berthière, and Benjamin Doyon for useful discussions. C.B thanks the National Defense of Canada for financial support to facilitate completion of his PhD. M.P. thanks the Perimeter Institute for hospitality where this work was begun.   We thank  NSERC of Canada for financial support.   Research at Perimeter Institute is supported by the Government of Canada through Industry Canada and by the Province of Ontario through the Ministry of Research and  Innovation.



\onecolumngrid

\section*{Additional Material}
\appendix
\section{BCHI model with arbitrary spin $s$}\label{A:BCHI}

The one dimensional BCHI model with $N\geq 2$ sites with the Hamiltonian, Eqn.\eqref{h} can be written as
\beq 
H_0=\frac{1}{2} \mathbf{S}^T A \mathbf{S}
\eeq 
where $\mathbf{S}^T=(S^z_1, S^z_2, \cdots, S^z_N)$ and $A$ is the circulant matrix with first row $(2a,b,0,\dots , 0,b)$. Because the Hamiltonian is quadratic and local, it can be defined on one-dimensional lattices of period 1 and 2. Specifically, the labels of the eigenstates $|s_1,\dots ,s_N\rangle$ should define a section $\sigma (i) = s_i$ on the orientable closed strip, the trivial fiber bundle $[-s,s]\times S^1$, on which the $z$ axis is sent to itself after one world trip. Alternatively, they could define a section on the Möbius strip, the (unique) non-orientable bundle that looks locally like the product $[-s,s]\times S^1$, on which the $z$ axis is sent to $-z$ after one world trip. Although the $z$ direction cannot be defined globally on the Möbius strip, it is defined locally so that nearest neighbour interactions $S^z_i S^z_{i+1}$ make sense and are independent of the local gauge $\pm z$. We will eventually find the phase diagram for all periodic chains. Since, in the thermodynamic limit, Landau's theorem precludes the existence of a phase transition in this system at any positive temperature, we study the phase diagram at $T=0$.

\subsection{The ground state on the orientable chain with $N$ even.}\label{S:GS_evenN}

On the orientable periodic chain with an even number of sites, the staggered spin operators $\bar S_j\equiv (-1)^j S_j$ are globally well-defined, and we find the usual mapping between the ferromagnetic ($b<0$) and the antiferromagnetic ($b>0$) cases
\beq
	H_0^{b>0} (\bar{S}_1, \dots , \bar{S}_N) = H_0^{b<0} (S_1, \dots , S_N)\label{st}
\eeq 
and vice versa.  For definiteness we will find the ferromagnetic ground state, and then obtain the antiferromagnetic one by the above duality. Note that staggered operators are locally defined when $N$ is odd, if not globally, and the equivalence Eqn.\eqref{st} remains locally valid.

The ferromagnetic case is easily dealt with.  We write the state $|s_1,\dots ,s_N \rangle$ as $|r(\alpha_1 , \dots , \alpha_N) \rangle$ where $r=\sqrt{\sum_k s_k^2}$ and $\sum_k \alpha_k^2=1$.  Then the energy is given as 
\begin{equation}\label{E:spherical}
E(s_1,\dots ,s_N)=\big(a+b\textstyle\sum_k \alpha_k \alpha_{k+1}\big) r^2 = C(\boldsymbol{\hat{\alpha}})r^2
\end{equation}
where explicitly, 
\beq
C(\boldsymbol{\hat{\alpha}})=\big(a+b\textstyle\sum_k \alpha_k \alpha_{k+1}\big) .\label{C}
\eeq
As $\sum_k(\alpha_k\pm \alpha_{k+1})^2\ge 0$, we have 
\beq
-1\leq\sum_k \alpha_k \alpha_{k+1} \leq 1.\label{bound}
\eeq  
Thus for $|b| < a$, using Eqn.\eqref{bound} in Eqn.\eqref{C}, we have $C(\boldsymbol{\hat{\alpha}})>0$ and hence the minimum energy configuration is realized exactly for $r=0$ corresponding to the state $|0,\dots ,0\rangle$ with corresponding energy $E_0=0$.  For half odd integer spin, the state $|0,\dots ,0\rangle$ is not permitted. Then in this case, one of the states closest to the origin, $|\pm 1/2,\dots ,\pm 1/2\rangle$ (with uncorrelated $\pm$ signs) will be the minimal energy configuration.  Since the first term in the energy does not care whether the spin is $\pm 1/2$ and since $b<0$, the energy is minimized at the ``little'' ferromagnetic  states 
\beq
 |1/2,\dots ,1/2\rangle \quad {\rm or}\quad   |-1/2,\dots ,-1/2\rangle\label{littleferro}
 \eeq  
with energy $E_0=(1/4)N^2(a-|b|)$.

For  all other cases,  $|b|>a$ (including $a$ negative), the factor $C(\boldsymbol{\hat{\alpha}})$ becomes negative for certain directions, and in particular for $\boldsymbol{\hat{\alpha}}^T = \pm \frac{1}{\sqrt{N}} (1,\dots , 1)$  the upper bound Eqn.\eqref{bound}, for the sum $\sum_k \alpha_k \alpha_{k+1}$ is saturated.  These are the only two states for which the bound is saturated, and here $C(\boldsymbol{\hat{\alpha}})=a+b=a-|b|$.   $C(\boldsymbol{\hat{\alpha}})$ is negative and minimal for this direction.   The extreme corners  $\pm(s,\dots ,s)$ of the hypercube $[-s,s]^N$ are attained along this direction.  Hence $r$ is maximal, and correspondingly, the energy is minimal at the two corners.  Thus the two corresponding ferromagnetic states, which we will write as  $ | \uparrow, \dots ,\uparrow \rangle$ and $ | \downarrow, \dots ,\downarrow \rangle$, are the ground states in the regime $|b|>a$ (and we will use the notation $\uparrow$ and $\downarrow$ when the corresponding spin is maximally up, $s$, or maximally down, $-s$, respectively).

When expressed as a function of the non-thermal parameter $a/|b|$ the ground state energy is non-analytic at $a=|b|$, so this is a quantum phase transition \cite{sachdev}. 
For $a=|b|$ the states   $| \pm (m, \dots ,m) \rangle$  are degenerate for any $m\in \{-s,-s+1,\dots ,s\}$ thus the ground state is $2s+1$ fold degenerate with $E_0=0$, and  the system passes through a highly degenerate critical point.  In the large $s$ limit this is veritably a massless continuum. 

\begin{figure}[ht]
\caption{Ground state of the orientable BCHI chain, with $N$ even, and arbitrary spin $s$}\label{F:Npair}
\includegraphics[]{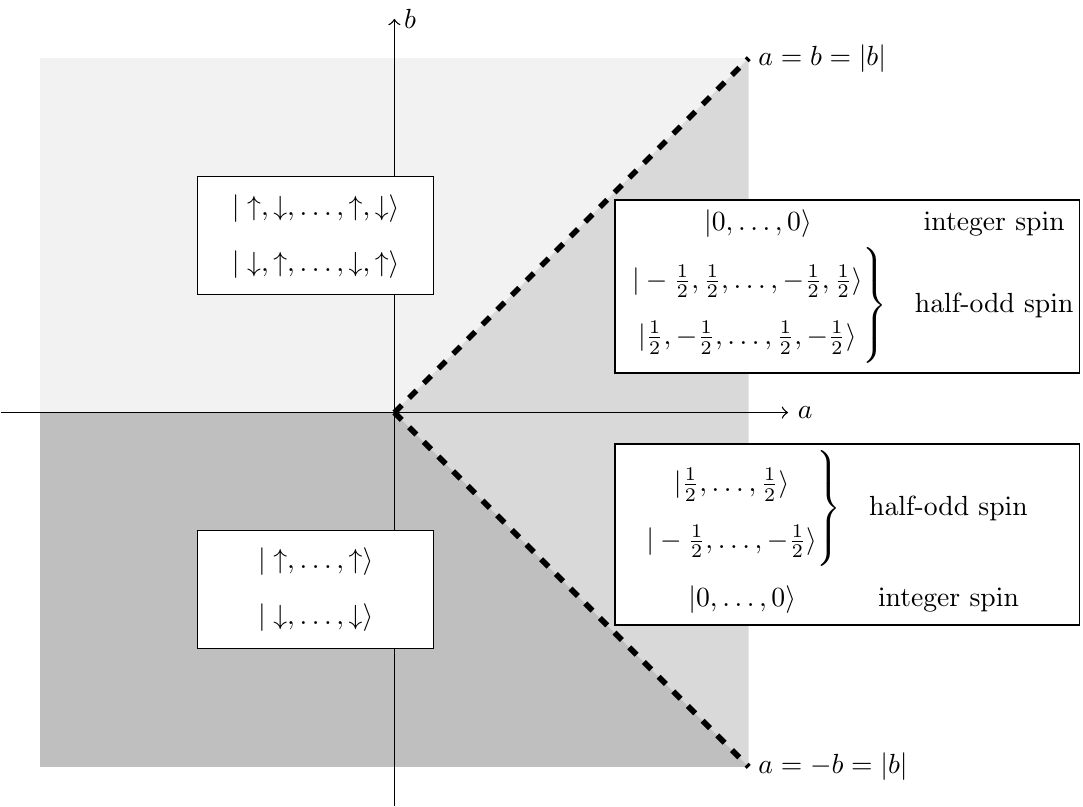}
\end{figure}

With the duality, Eqn.\eqref{st}, we obtain the antiferromagnetic ground state, and plot the result in Fig.\eqref{F:Npair}. The ground state is everywhere doubly degenerate, except on the critical lines $a=|b|$, as well as for the particular case $a>|b|$ and integer spin $s$, for which the ground state 
$|0,\dots ,0\rangle$ trivially possesses the $\mathbb{Z}_2$ symmetry of the Hamiltonian. We now find, for all values of $a$ and $b$, the soliton interpolating between the two degenerate vacua, if applicable.

\subsection{The BCHI soliton}\label{S:BCHIsoliton}
The profile of the soliton will be seen to depend only on the ratio $a/|b|$, and not on the size $N$ of the lattice nor on the boundary conditions. The calculation and results are local, and will apply equally well to any periodic lattice, whether or not the degenerate ground states identified in Fig.\eqref{F:Npair} can be realized globally. For definiteness, let us begin with the antiferromagnetic case ($b>0$). 

\subsubsection{Antiferromagnetic soliton}\label{S:AF}
\noindent When $0<b<a$, a straightforward induction shows that the soliton (for half-odd spin $s$) is the ``little" defect comprising two adjacent $+1/2$ or two adjacent $-1/2$ spin components. 

For $0<a<b$, the soliton is found by minimizing the energy functional subject to the boundary conditions, $s_0 = s$ and $s_{n+1} = (-1)^n s$, connecting the two Néel ground states. For now $n$ is at least as large as the size of the soliton, but otherwise arbitrary. With these boundary conditions, the energy of the generic state $|s_0,\dots , s_{n+1}\rangle$ is
\beq
E=\frac{1}{2}\mathbf{s}^{\text{T}} B_n \mathbf{s} + 2as^2 + b\mathbf{s}^{\text{T}}\mathbf{t}\; ,
\eeq
where $B_{n}$ is the tridiagonal Toeplitz matrix of dimension $n\times n$ with the three non-zero diagonals given by
\beq\label{B}
	B_{n}=\left(\begin{matrix}

		 2a& b & &  &  &  &&  \\

		b&2a  & b &  &  &  & &\\

		 & b &2a  & b &  &  & & \\
		
		 &&\cdot&\cdot&\cdot&&&\\
		 		 
		 &&&\cdot&\cdot&\cdot&&\\

		 &&&&\cdot&\cdot&\cdot&\\
		 				
&&&&&b&2a&b\\
		 &&&&&&b&2a\\
	\end{matrix}
	\right)_{n\times n}
\eeq
and where $\mathbf s = (s_1, s_2, \dots , s_{n})^T$ and $\mathbf t = (s,0,\dots ,0,(-1)^n s)^T$. The critical points of the energy form are given by
\beq\label{E:crit1}
B_{n}\mathbf s=-b\mathbf t.
\eeq 
The Hessian matrix of the energy quadratic form is $B_n$, whose eigenvalues and eigenvectors are easily found.  The eigenvalues are $\lambda_k = 2\big(a+b\cos \frac{k\pi}{n+1}\big)$, $k=1,2, \dots , n$.  Now the Hessian is positive definite for $\cos\frac{\pi}{n+1}<\frac{a}{b}$, therefore in this range the unique critical point, Eqn. \eqref{E:crit1}, is the minimum energy configuration for the boundary problem $s_0 = s$, $s_{n+1} = (-1)^n s$.  The solution is obtained by inverting the Toeplitz matrix $B_{n}$, $\mathbf s=-b(B_{n})^{-1}\mathbf t$, explicitly from \cite{kst,fp}
\begin{equation}\label{E:sk1}
	s_k = \bar{s}^{\, n}_k \overset{\rm def}{=} (-1)^k s \bigg( \frac{\sin (n+1-k)\theta - \sin k\theta}{\sin (n+1)\theta}\bigg),\qquad\qquad k=1,\dots , n,
\end{equation}
where $\cos \theta = a/b$.   Equivalently, the solution can be written as
\begin{equation}\label{E:sk2}
	\bar{s}^{\, n}_k=\frac{(-1)^k s}{\sin (n+1)\theta / 2} \;\sin \big(\tfrac{n+1}{2} - k\big)\theta, \qquad\qquad k=1,\dots ,n.
\end{equation}
From this expression it is easily seen that $|\bar{s}^{\, n}_k| < s$, so the solution is inside the hypercube $[-s,s]^n$ of acceptable solutions. One also recognizes a rotation of the spin components by $\pi$, interpolating smoothly from one Néel ground state to the other over the sites labelled by $k=1,\dots , n$. The soliton we are after must therefore be found among these functions $\bar{s}^{\, n}_k$, with integer $n$ such that $\cos\frac{\pi}{n+1}<\frac{a}{b}<1$, since they are the minimum energy solutions to the boundary problems $s_0 = s$, $s_{n+1} = (-1)^n s$. It is clear that the corresponding energies decrease with $n$, $E_n>E_{n+1}$, since the $n$th problem is subsumed in the $(n+1)$st. This implies that the soliton has maximal such $n$. We conclude that the antiferromagnetic ($b>0$) soliton for $a>0$ is given by Eqn.\eqref{E:sk2}:
\beq\label{E:skm}
s_k = \bar{s}^{\, M}_k  = \frac{(-1)^k s}{\sin (M+1)\theta / 2} \;\sin \big(\tfrac{M+1}{2} - k\big)\theta,\qquad\qquad k=1,\dots , M,
\eeq
where $\cos\theta = \frac{a}{b}$ and $M\geq 1$ is the unique integer such that $\cos\frac{\pi}{M+1}<\frac{a}{b}<\cos\frac{\pi}{M+2}$.  The soliton corresponds to a rotation of the spin components by $\pi$ over the sites labelled by $k=1,\dots , M$, interpolating smoothly from one Néel ground state to the other, and has minimal energy among such interpolations. As $a$ is decreased towards zero, the soliton is shortened until it reaches the trivial $\bar{s}^1_1 = 0$ in the range $0<a<\cos\frac{\pi}{3}$. 

As $a$ becomes negative one can guess, and prove by an easy induction, that the domain-wall soliton collapses to a simple ``up-up" or ``down-down" defect. 

\subsubsection{Ferromagnetic soliton}\label{S:F}
In the ferromagnetic case $b<0$, the problem can be solved by an essentially identical analysis or more simply by the exact local equivalence, Eqn.\eqref{st}. In places where the antiferromagnetic case had ``up-up" or ``down-down" defects (or little defects), the ferromagnetic case has ``up-down"  or ``down-up" defects (or little defects). Where the antiferromagnetic ground state is non-degenerate, so is the ferromagnetic one. Where the antiferromagnetic soliton is $s_k = \bar{s}^M_k$, Eqn.\eqref{E:skm}, the ferromagnetic soliton is
\beq\label{E:sk3}
s_k = s^M_k = \frac{s}{\sin (M+1)\theta / 2} \;\sin \big(\tfrac{M+1}{2} - k\big)\theta,\qquad\qquad k=1,\dots , M,
\eeq
with $\cos\theta = \frac{a}{|b|}$ and $M\geq 1$ is the unique integer such that $\cos\frac{\pi}{M+1}<\frac{a}{|b|}<\cos\frac{\pi}{M+2}$. 

\subsubsection{BCHI soliton in parameter space}
\begin{figure}[ht]
\caption{Soliton in parameter space}\label{F:domain-wall}
\includegraphics{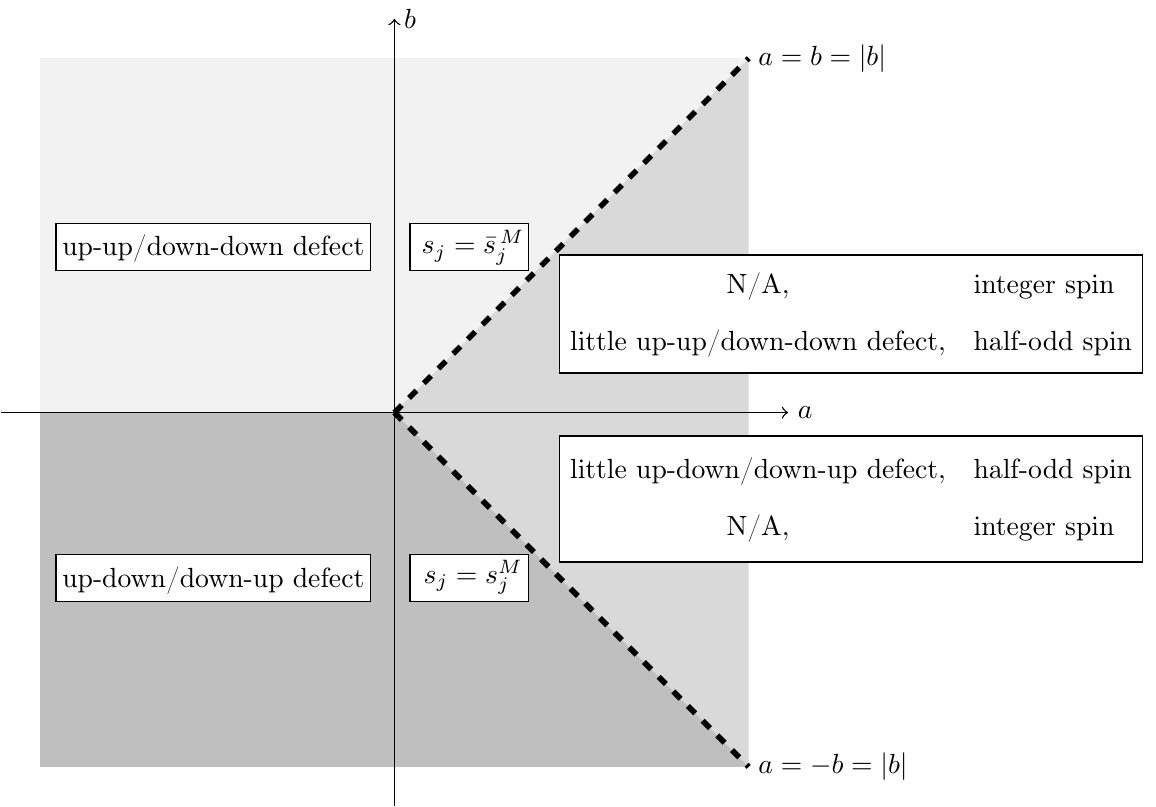}
\end{figure}

These results are summarized in Fig.~\ref{F:domain-wall}, where defects are made of spin components $\pm s$, ``little'' defects are made of spin components $\pm \frac{1}{2}$, while $s^M_j$ and $\bar s^{\, M}_j$ are solitons of length $M\geq 1$ in the ferromagnetic ($b<0$) and antiferromagnetic ($b>0$) regions, respectively. 
From Eqns.\eqref{E:sk2} and~\eqref{E:sk3}, their profile is
\beq\label{E:sol_antisol}
s_j = \begin{cases}
s^M_j &= \frac{s}{\sin (M+1)\theta / 2} \;\sin \big(\tfrac{M+1}{2} - j\big)\theta,\qquad\qquad b<0 \text{ (ferromagnetic)},\\
\bar{s}^{\,M}_j &=\frac{(-1)^j s}{\sin (M+1)\theta / 2} \;\sin \big(\tfrac{M+1}{2} - j\big)\theta , \qquad\qquad b>0 \text{ (antiferromagnetic)},
\end{cases}
\eeq
where $\cos\theta = \frac{a}{|b|}$, and $M\geq 1$ is the unique integer such that $\cos\frac{\pi}{M+1}<\frac{a}{|b|}<\cos\frac{\pi}{M+2}$.

 \subsubsection{Explict, discrete-spin soliton profile}\label{S:discrete_soliton_appendix}
 The expressions found above in Eqns.(\ref{E:skm}, \ref{E:sk3}) for the soliton profile are expressed as continuous functions of the parameters and hence do not actually correspond to the discrete values that are allowed for the $z$ components of the spin.  However it is an easy numerical exercise to find the actual discrete soliton profiles.  Using \textsc{Mathematica} we find a few examples for the antiferromagnetic case. See Tables~\ref{Spin 1},~\ref{Spin 2},~\ref{Spin 3},~\ref{Spin 3/2},~\ref{Spin 5/2},~\ref{Spin 7/2}. For small values of the spin we observe that the exact, discrete soliton changes size at rational values of $\frac{a}{b}\sim\frac{n}{n+1}$. At the present time we only have a numerical observation of this phenomenon, and we think this is only for small values of the spin, as it no longer seems to be the case already at spin 20 and spin 41/2. (See Figs.\ref{F:Spin20},~\ref{F:Spin41half} of the main text.) For these values, the soliton's length is numerically observed to change from $M$ to $M+1$ at $a/b \sim \cos\pi/(M+2)$, in agreement with the large spin calculation.

\begin{table}[p]
\caption{Spin 1}\label{Spin 1}	
		\begin{tabular}{ | c | c | }\hline
\multicolumn{2}{| c |}{$\uparrow\leftrightarrow s_z=1;\quad 0\leftrightarrow s_z=0$} \\\hline
$\left| \dots,\downarrow,\uparrow, 0,   \downarrow,\uparrow,\dots \right\rangle$&$\left| \dots ,0, 0, 0, 0, 0,\dots \right\rangle$\\ \hline
$\frac{a}{b}< 1$&$\frac{a}{b}>1$\\ \hline
		\end{tabular}
\end{table}
\begin{table}[p]
\caption{Spin 2}\label{Spin 2}	
		\begin{tabular}{ | c | c | c | c | }\hline
\multicolumn{4}{ |c| }{$\big\uparrow\leftrightarrow s_z=2;\quad\uparrow\leftrightarrow s_z=1;\quad 0\leftrightarrow s_z=0$} \\ \hline
$\left| \dots,\big\downarrow,\big\uparrow, 0,   \big\downarrow,\big\uparrow,\dots \right\rangle$&$\left|  \dots,\big\downarrow,\big\uparrow, \downarrow,\downarrow,   \big\uparrow,\big\downarrow,\dots \right\rangle $&$\left| \dots,\big\downarrow,\big\uparrow, \downarrow,0,\uparrow,   \big\downarrow,\big\uparrow,\dots \right\rangle$&$\left| \dots ,0, 0, 0, 0, 0,\dots \right\rangle$\\ \hline
$\frac{a}{b}< \haf 1$&$\haf 1<\frac{a}{b}<\frac{3}{4}$&$\frac{3}{4}<\frac{a}{b}<1$&$\frac{a}{b}>1$\\ \hline
		\end{tabular}
	\end{table}
\begin{table}[p]
\caption{Spin 3}\label{Spin 3}	
		\resizebox{7in}{!}{%
		\begin{tabular}{ | c | c | c | c | c | c | }\hline
\multicolumn{6}{ |c| }{$\Big\uparrow\leftrightarrow s_z=3;\quad\quad\big\uparrow\leftrightarrow s_z=2;\quad\quad\uparrow\leftrightarrow s_z=1;\quad 0\leftrightarrow s_z=0$} \\ \hline
\multirow{2}{*}{$\left| \dots,\Big\downarrow,\Big\uparrow, 0,   \Big\downarrow,\Big\uparrow,\dots \right\rangle$}&$\left| \dots,\Big\downarrow,\Big\uparrow,  \big\downarrow,\downarrow,\Big\uparrow,\Big\downarrow,\dots \right\rangle$&\multirow{2}{*}{$\left| \dots,\Big\downarrow,\Big\uparrow,\big\downarrow, 0, \big\uparrow,  \Big\downarrow,\Big\uparrow,\dots \right\rangle$}&\multirow{2}{*}{$\left| \dots,\Big\downarrow,\Big\uparrow,\big\downarrow, \uparrow,\uparrow, \big\downarrow,  \Big\uparrow,\Big\downarrow,\dots \right\rangle$}&\multirow{2}{*}{$\left| \dots,\Big\downarrow,\Big\uparrow,\big\downarrow, \uparrow,0,\downarrow, \big\uparrow,  \Big\downarrow,\Big\uparrow,\dots \right\rangle$}&\multirow{2}{*}{$\left| \dots ,0, 0, 0,\dots \right\rangle$}\\ 
\cline{2-2}
&$\left| \dots,\Big\downarrow,\Big\uparrow,  \downarrow,\big\downarrow,\Big\uparrow,\Big\downarrow,\dots \right\rangle$&&&&\\ \hline
$\frac{a}{b}< \haf 1$&$\haf 1<\frac{a}{b}< \frac{2}{3}$&$\frac{2}{3}<\frac{a}{b}< \frac{6}{7}$&$\frac{6}{7}<\frac{a}{b}< \frac{8}{9}$&$\frac{8}{9}<\frac{a}{b}<1$&$\frac{a}{b}>1$\\ \hline
		\end{tabular}%
		}
		\end{table}		
\begin{table}[p]
\caption{Spin $\frac{3}{2}$}\label{Spin 3/2}	
		\begin{tabular}{ | c | c | c | }\hline
\multicolumn{3}{ |c| }{$\big\uparrow\leftrightarrow s_z=\frac{3}{2};\quad\uparrow\leftrightarrow s_z=\frac{1}{2}$} \\ \hline
$\left| \dots,\big\downarrow,\big\uparrow, \uparrow,   \big\downarrow,\big\uparrow,\dots \right\rangle$&$\left|  \dots,\big\downarrow,\big\uparrow, \downarrow,\downarrow,   \big\uparrow,\big\downarrow,\dots \right\rangle $&$\left| \dots,\downarrow,\uparrow, \downarrow,\downarrow,\uparrow,   \downarrow,\uparrow,\dots \right\rangle$\\ \hline
$\frac{a}{b}< \haf 1$&$\haf 1<\frac{a}{b}<1$&$\frac{a}{b}>1$\\ \hline
		\end{tabular}
\end{table}
\begin{table}[p]
\caption{Spin $\haf 5$}\label{Spin 5/2}	
	\resizebox{7in}{!}{
		\begin{tabular}{ | c | c | c | c | c | }\hline
\multicolumn{5}{ |c| }{$\Big\uparrow\leftrightarrow s_z=\frac{5}{2};\quad\big\uparrow\leftrightarrow s_z=\frac{3}{2};\quad\uparrow\leftrightarrow s_z=\frac{1}{2}$} \\ \hline
\multirow{2}{*}{$\left| \dots,\Big\downarrow,\Big\uparrow, \uparrow,   \Big\downarrow,\Big\uparrow,\dots \right\rangle$}&$\left| \dots,\Big\downarrow,\Big\uparrow,  \big\downarrow,\downarrow,\Big\uparrow\Big\downarrow,\dots \right\rangle$&\multirow{2}{*}{$\left| \dots,\Big\downarrow,\Big\uparrow,\big\downarrow, \uparrow, \big\uparrow,  \Big\downarrow,\Big\uparrow,\dots \right\rangle$}&\multirow{2}{*}{$\left| \dots,\Big\downarrow,\Big\uparrow,\big\downarrow, \uparrow,\uparrow, \big\downarrow,  \Big\uparrow,\Big\downarrow,\dots \right\rangle$}&\multirow{2}{*}{$\left| \dots,\downarrow,\uparrow, \downarrow,\downarrow,\uparrow,   \downarrow,\uparrow,\dots \right\rangle$}\\ 
\cline{2-2}
&$\left| \dots,\Big\downarrow,\Big\uparrow,\downarrow,\big\downarrow,\Big\uparrow,\Big\downarrow,\dots \right\rangle$&&&\\ \hline
$\frac{a}{b}< \haf 1$&$\haf 1<\frac{a}{b}< \frac{3}{4}$&$\frac{3}{4}<\frac{a}{b}< \frac{5}{6}$&$\frac{5}{6}<\frac{a}{b}< 1$&$\frac{a}{b}>1$\\ \hline
		\end{tabular}
		}
\end{table}
\begin{table}[p]
\caption{Spin $\haf 7$}\label{Spin 7/2}	
\resizebox{7in}{!}{%
		\begin{tabular}{ | c | c | c | c | c | c | c | }\hline
\multicolumn{7}{ |c| }{$\bigg\uparrow\leftrightarrow s_z=\haf 7;\quad\quad\Big\uparrow\leftrightarrow s_z=\haf 5;\quad\quad\big\uparrow\leftrightarrow s_z=\haf 3;\quad\quad\uparrow\leftrightarrow s_z= \haf 1;\quad 0\leftrightarrow s_z=0$} \\ \hline
\multirow{2}{*}{$\left| \dots,\bigg\downarrow,\bigg\uparrow, \uparrow,   \bigg\downarrow,\bigg\uparrow,\dots \right\rangle$}&\multirow{2}{*}{$\left| \dots,\bigg\downarrow,\bigg\uparrow, \big\downarrow,\big\downarrow,   \bigg\uparrow,\bigg\downarrow,\dots \right\rangle$}&$\left| \dots,\bigg\downarrow,\bigg\uparrow, \Big\downarrow,\uparrow,\big\uparrow,   \bigg\downarrow,\bigg\uparrow,\dots \right\rangle$&$\left| \dots,\bigg\downarrow,\bigg\uparrow, \Big\downarrow,\uparrow,\big\uparrow,\Big\downarrow,   \bigg\uparrow,\bigg\downarrow,\dots \right\rangle$&\multirow{2}{*}{$\left| \dots,\bigg\downarrow,\bigg\uparrow, \Big\downarrow,\big\uparrow,\uparrow,\big\downarrow,\Big\uparrow,   \bigg\downarrow,\bigg\uparrow,\dots \right\rangle$}&\multirow{2}{*}{$\left| \dots,\bigg\downarrow,\bigg\uparrow, \Big\downarrow,\big\uparrow,\downarrow,\downarrow,\big\uparrow,\Big\downarrow,   \bigg\uparrow,\bigg\downarrow,\dots \right\rangle$}&\multirow{2}{*}{$\left| \dots ,0, 0, 0,\dots \right\rangle$}\\ 
\cline{3-4}
&&$\left| \dots,\bigg\downarrow,\bigg\uparrow, \big\downarrow,\downarrow,\Big\uparrow,   \bigg\downarrow,\bigg\uparrow,\dots \right\rangle$&$\left| \dots,\bigg\downarrow,\bigg\uparrow, \Big\downarrow,\big\uparrow,\uparrow,\Big\downarrow,   \bigg\uparrow,\bigg\downarrow,\dots \right\rangle$&&&\\ \hline
$\frac{a}{b}< \haf 1$&$\haf 1<\frac{a}{b}< \frac{3}{4}$&$\frac{3}{4}<\frac{a}{b}< \frac{5}{6}$&$\frac{5}{6}<\frac{a}{b}< \frac{9}{10}$&$\frac{9}{10}<\frac{a}{b}<\frac{11}{12}$&$\frac{11}{12}<\frac{a}{b}<1$&$\frac{a}{b}>1$\\ \hline
		\end{tabular}%
		}
\end{table}

\section{Solitons of various sizes with the perturbative Heisenberg term\label{D}}
\subsection{Solitons of length one}\label{S:length_one}
We now consider the full Hamiltonian, Eqn.\eqref{E:H_ab}, and treat the Heisenberg term perturbatively. For now, we specialize the the region $0<a<\frac{b}{2}$ of parameter space, where the classical soliton has size one (one non-maximal spin $z$ component). This case deserves special treatment because small-soliton translations by one lattice constant occur at low perturbative order only for solitons of length one. For other intermediate soliton lengths, translations by \textit{two} lattice constants occur earlier in the perturbative development than translations by one lattice constant. (See Sec. \ref{S:any_length}.) We compute the perturbative energy splitting in the soliton degenerate subspace due to translations by one lattice constant on the periodic chain of odd length $N$. 

\subsubsection{Integer spin}
When $s\in\mathbb{N}$, the soliton has total spin zero, and the soliton degenerate subspace has dimension $2N$. For each soliton $|\nu\rangle$, where $\nu$ indicates position on the chain, define the projector $P_{\nu}=\sum_{\mu\neq\nu} |\mu\rangle\langle\mu|$. Let $|1\rangle$ have $z$ component representation
\beq
|\dots, \uparrow,\downarrow,\uparrow , \underbrace{0, \downarrow}_{\mathclap{\text{sites 1 and 2}}},\uparrow,\downarrow,\dots\rangle,
\eeq 
where arrows denote maximal components $\pm s$. (The soliton with Néel background reversed, denoted $|\overline{1}\rangle$, is also one of the $|\nu\rangle$'s.) Applying $P_1(S_1^- S_2^+ + S_1^+ S_2^-)^s$ to $|1\rangle$ will produce the state with $z$ component representation
\beq
|\dots, \uparrow,\downarrow,\uparrow , \underbrace{\downarrow, 0}_{\mathclap{\text{sites 1 and 2}}},\uparrow,\downarrow,\dots\rangle.
\eeq
It seems natural to denote it $|2\rangle$. It is clear that the transition $|1\rangle\to|2\rangle$ occurs at perturbative order $s$, and not earlier, because a minimal amount $s$ of spin must be exchanged between sites 1 and 2, no matter what. Note also that $P_2(S_1^- S_2^+ + S_1^+ S_2^-)^s$ will operate the inverse transition $|2\rangle\to|1\rangle$ when applied on $|2\rangle$. The soliton $|\mu+1\rangle$ is defined recursively as
\beq\label{E:TOp}
|\mu+1\rangle \propto P_{\mu}(S_{\mu}^- S_{\mu+1}^+ + S_{\mu}^+ S_{\mu+1}^-)^s |\mu\rangle,
\eeq
where $S_{\mu}^- S_{\mu+1}^+ + S_{\mu}^+ S_{\mu+1}^-$ is shorthand for $S_{(\mu\text{ mod }N)}^- S_{(\mu+1\text{ mod }N)}^+ + S_{(\mu\text{ mod }N)}^+ S_{(\mu+1\text{ mod }N)}^-$.
One easily convinces oneself that $|N+1\rangle$ thus defined is actually $|\overline{1}\rangle$, and that $|2N+1\rangle = |\overline{N+1}\rangle = |1\rangle$. All translations are reversible at order $s$, since 
\beq\label{E:TOp-}
|\mu \rangle \propto P_{\mu + 1}(S_{\mu}^- S_{\mu+1}^+ + S_{\mu}^+ S_{\mu+1}^-)^s |\mu + 1\rangle.
\eeq 
Translations over more than one lattice constant will require more adjacent transpositions of spin components, and will occur at higher orders. Brillouin-Wigner perturbation theory tells us that the level splitting is given to lowest order by the eigenvalues of the matrix $w$ with (off-diagonal) components
\beq\label{E:b}
w_{\mu\nu} = \langle\mu | V\left(R_{\nu}V\right)^{s-1}|\nu\rangle,\hspace{40pt}\mu,\nu = 1,2,\dots, 2N,
\eeq
where
\beq\label{E:Rv}
R_{\nu}=(E_{\nu}-H_0)^{-1}Q = (E_{\nu}-H_0)^{-1}\left(1-\sum_{\mu} |\mu\rangle\langle\mu|\right),
\eeq
and $E_{\nu}$ is the exact energy of the perturbed soliton of quantum number $\nu$. Translation invariance of the translation operators from Eqns. \eqref{E:TOp}-\eqref{E:TOp-} implies that $w$ is a circulant matrix,
\beq
w=\text{circ}(a_1,\dots,a_{2N})=\left(\begin{matrix}
	a_1 & a_2 & \cdots & a_{2N}\\
	a_{2N} & a_1 & \cdots & a_{2N-1}\\
	\vdots & \vdots & \ddots & \vdots\\
	a_2 & a_3 & \cdots & a_1
	\end{matrix}\right). 
\eeq
It is clear that $w$ is symmetric, and that its first column has only two nonzero (off-diagonal) entries, $w_{2,1}$ and $w_{2N,1}$, which must be equal by the combination of circulation and symmetry. So it suffices to find the constant $C$ such that $w=C\,\text{circ}(0,1,0,\dots , 0,1)$. We have
\beq\label{E:C_initial}
C=w_{2,1}=\left(\frac{|J|}{2}\right)^s \langle 2|S_{1}^- S_{2}^+ \left(\tfrac{Q}{E_1 - H_0}\, S_{1}^- S_{2}^+ \right)^{s-1}|1\rangle
= \left(\frac{|J|}{2}\right)^s \frac{\langle 2|(S_{1}^- S_{2}^+ )^s |1\rangle}{\prod_{m=1}^{s-1}(E_1 - e_m)},
\eeq
where $e_m$ is the unperturbed energy of the state
\beq
(S_{1}^- S_{2}^+ )^m |1\rangle \propto |\dots, \uparrow,\downarrow,\uparrow , \underbrace{-m, -s+m}_{\mathclap{\text{sites 1 and 2}}},\uparrow,\downarrow,\dots\rangle.
\eeq
Approximating $E_1$ as $E_1 (|J|=0)$, the energy denominators are easily computed to be
\beq\label{E:en_denom}
E_1 - e_m = (2a-b)(s-m)m.
\eeq
The amplitude $\langle 2|(S_{1}^- S_{2}^+ )^s |1\rangle$ in \eqref{E:C_initial} is obtained by inserting identity resolutions
\beq
\begin{aligned}
\langle 2|(S_{1}^- S_{2}^+ )^s |1\rangle&=\prod_{m=0}^{s-1} \langle\dots, \uparrow,\downarrow,\uparrow , \underbrace{-m-1, -s+m+1}_{\mathclap{\text{sites 1 and 2}}},\uparrow,\downarrow,\dots | S_{1}^- S_{2}^+  
|\dots, \uparrow,\downarrow,\uparrow , \underbrace{-m, -s+m}_{\mathclap{\text{sites 1 and 2}}},\uparrow,\downarrow,\dots\rangle\\
&= \prod_{m=0}^{s-1} \sqrt{(s-m)(s+1+m)(2s-m)(m+1)}\\
&= \prod_{m=0}^{s-1} (2s-m)(s-m).
\end{aligned}
\eeq
Hence
\beq\label{E:C}
C=\left(\frac{|J|}{2}\right)^s \frac{\prod_{m=0}^{s-1} (2s-m)(s-m)}{\prod_{m=1}^{s-1}(2a-b)(s-m)m} = (-1)^{s-1}Ks^2 \frac{(2s)!}{(s!)^2}\left(\frac{|J|}{2K}\right)^s ,
\eeq
where we have introduced $K = b-2a >0$. The factor $(-1)^{s-1}$ is due to the presence of $s-1$ negative energy denominators. Circulant matrices $\text{circ}(a_1 , \dots, a_{2N})$ are diagonalized by discrete Fourier transforms, and have eigenvalues
\beq
\epsilon_j = a_1 + a_2 \omega_j + a_3 \omega_j^2 + \dots + a_{2N}\omega_j^{2N-1},\hspace{40pt} j=0,\dots, 2N-1,
\eeq
and corresponding eigenvectors $\frac{1}{\sqrt{2N}}(1,\omega_j,\omega_j^2,\dots ,\omega_j^{2N-1})^{\text{T}}$, with $\omega_j = \exp(ij\pi /N)$. As the matrix $w$ from \eqref{E:b} has the form $w=C\,\text{circ}(0,1,0,\dots , 0,1)$, its eigenvalues are
\beq\label{E:epsilon_length_one}
\epsilon_j = C(\omega_j + \omega_j^{2N-1})= 2C \cos \frac{j\pi}{N},\hspace{40pt} j=0,\dots, 2N-1.
\eeq
The spectrum is doubly degenerate for all values of $j$ except $j=0$ and $j=N$. From \eqref{E:C} the ground state of the periodic chain corresponds to $j=0$ when $s$ is even, and to $j=N$ when $s$ is odd:
\beq\label{E:gs_evenspin}
|\psi_0\rangle = \begin{cases}
	\frac{1}{\sqrt{2N}}\sum_{\mu}|\mu\rangle\hspace{10pt}&,\hspace{20pt}s\text{ even};\\
	\frac{1}{\sqrt{2N}}\sum_{\mu}(-1)^{\mu}|\mu\rangle\hspace{10pt}&,\hspace{20pt}s\text{ odd}.
\end{cases}
\eeq
We remark that when $a\to b/2$, at the crossover with solitons of length two, we get $K\to 0$, and our expression for $C$, Eqn.\eqref{E:C}, will diverge. Thus it appears that Brillouin-Wigner perturbation theory could fail to converge in that limit, due to the presence of small energy denominators in \eqref{E:C_initial}. But refining our expression for the energy denominators \eqref{E:en_denom} by using the semiclassical energy (including the Heisenberg term) of spin coherent states, we get instead
\beq
E_1 - e_m = (2a-b-|J|)(s-m)m,
\eeq
and $K=b+|J| - 2a$ in Eqn.\eqref{E:C}. At the crossover $a\to b/2$, we obtain $K\to |J|$, and find that $C$ is of order $|J|$, so that perturbative theory should converge after all. 

\subsubsection{Half-odd spin}\label{S:half-odd}
When $s-\tfrac{1}{2}\in\mathbb{N}$, the soliton has total spin $\pm\frac{1}{2}$, and the soliton degenerate subspace has dimension $4N$. To compute the perturbative level splitting due to soliton translations, we need to diagonalize a $4N\times 4N$ Brillouin-Wigner matrix $W$, analogous to the matrix $w$ from the previous section, Eqn.\eqref{E:b}. But since the Hamiltonian conserves total spin, that matrix is block diagonal
\beq
W=\left(\begin{matrix}
	w^+ & 0\hspace{4pt}\\
	0\hspace{5pt} & w^-\end{matrix}\right),
\eeq
and we need only to diagonalize each $2N\times 2N$ superselection sector separately. Define the projectors $P_{\nu}^{\pm}=\sum_{\mu\neq\nu} {|\mu\rangle}_{\pm}\prescript{}{\pm}{\langle\mu|}$ into the degenerate soliton subspaces of total spin $\pm\frac{1}{2}$, respectively, and let 
\beq
{|1\rangle}_{\pm} = |\dots, \uparrow,\downarrow,\uparrow , \underbrace{\pm\tfrac{1}{2}, \downarrow}_{\mathclap{\text{sites 1 and 2}}},\uparrow,\downarrow,\dots\rangle.
\eeq 
(The solitons with all components reversed, denoted ${|\overline{1}\rangle}_{\pm}$, are part of the ${|\nu\rangle}_{\mp}$'s.) Now
\beq
P_{1}^- (S_1^- S_2^+ + S_1^+ S_2^-)^{s-\frac{1}{2}}{|1\rangle}_{-}\quad\propto\quad |\dots, \uparrow,\downarrow,\uparrow , \underbrace{\downarrow, -\tfrac{1}{2}}_{\mathclap{\text{sites 1 and 2}}},\uparrow,\downarrow,\dots\rangle.
\eeq
Like in the previous section, we call the righthand side ${|2\rangle}_{-}$. The transition ${|1\rangle}_{-}\to{|2\rangle}_{-}$ and its inverse occur at perturbative order $s-\frac{1}{2}$, and not earlier. The transition ${|2\rangle}_{-}\to{|3\rangle}_{-}$, however, occurs only at the \textit{next} order in perturbation, $s+\frac{1}{2}$. Indeed, transposing spin components $s_2=-\frac{1}{2}$ and $s_3=+s$ requires at least $|s_2 - s_3 |$ instances of 
$S_2^- S_3^+ + S_2^+ S_3^-$. At minimal perturbative order $s-\frac{1}{2}$, the spin $-\frac{1}{2}$ sector only has translations of the type
\beq\label{E:trans_minus}
{|2k-1\rangle}_{-} \longleftrightarrow {|2k\rangle}_{-} \hspace{30pt},\hspace{30pt} k=1,2,\dots,N.
\eeq
As before, ${|N+1\rangle}_{-}={|\overline{1}\rangle}_{-}$. The Brillouin-Wigner matrix for this sector of soliton subspace has components 
\beq
w_{\mu\nu}^{-} = \prescript{}{-}{\langle}\mu | V\left(R_{\nu}^{-}V\right)^{s-\frac{3}{2}}{|\nu\rangle}_{-}\hspace{30pt} ,\hspace{40pt}\mu,\nu = 1,2,\dots, 2N,
\eeq
where
\beq
R_{\nu}^{-}=(E_{\nu}^{-}-H_0)^{-1}Q_{-} = (E_{\nu}^{-}-H_0)^{-1}\left(1-\sum_{\mu} {|\mu\rangle}_{-}\prescript{}{-}{\langle\mu|}\right),
\eeq
and $E_{\nu}^{-}$ is the exact energy of the perturbed soliton of quantum number $\nu$ and total spin $-\tfrac{1}{2}$. According to \eqref{E:trans_minus}, the matrix $w^-$ is block diagonal, with $N$ blocks of dimension $2\times 2$ along the diagonal. The blocks are symmetric, and they must be identical by the two-site translation invariance of the Néel background. What we need to diagonalize is thus a $2\times 2$ matrix
$C^- \left(\begin{smallmatrix}
	0 &  1\\
	1 & 0\end{smallmatrix}\right),$
with eigenvalues 
\beq
\epsilon_{\pm} = \pm C^- ,
\eeq
and corresponding eigenvectors $\frac{1}{\sqrt{2}}(1,\pm1)^{\text{T}}$, respectively. At this minimal perturbative order, the spectrum of the spin $-\frac{1}{2}$ sector is $N$-fold degenerate, and has a mass gap of $2|C^-|$. The constant $C^-$ is
\beq\label{E:C_minus}
C^- = w^{-}_{2,1}=\left(\frac{|J|}{2}\right)^{s-\frac{1}{2}} \prescript{}{-}{\langle 2|}S_{1}^- S_{2}^+ \left(\tfrac{Q}{E_1^- - H_0}\, S_{1}^- S_{2}^+ \right)^{s-\frac{3}{2}}
{|1\rangle}_{-}
= \left(\frac{|J|}{2}\right)^{s-\frac{1}{2}} \frac{\prescript{}{-}{\langle 2|}(S_{1}^- S_{2}^+ )^{s-1/2} {|1\rangle}_{-}}{\prod_{m=1}^{s-3/2}(E_1^- - e_m)},
\eeq
where 
\beq
E_1^- - e_m = (2a-b)(s-m-\tfrac{1}{2})m.
\eeq
The amplitude $\prescript{}{-}{\langle 2|}(S_{1}^- S_{2}^+ )^{s-\frac{1}{2}} {|1\rangle}_{-}$ in \eqref{E:C_minus} is obtained by introducing identity resolutions, and the result is 
\beq
\begin{aligned}
\prescript{}{-}{\langle 2|}(S_{1}^- S_{2}^+ )^{s-\frac{1}{2}} {|1\rangle}_{-} &= \prod_{m=0}^{s-\frac{3}{2}} \sqrt{(s-m-\tfrac{1}{2})(s+m+\tfrac{3}{2})(2s-m)(m+1)}\\
&= \prod_{m=0}^{s-\frac{3}{2}} (2s-m)(s-m-\tfrac{1}{2}).
\end{aligned}
\eeq
Hence
\beq
C^- =\left(\frac{|J|}{2}\right)^{s-\frac{1}{2}} \frac{\prod_{m=0}^{s-\frac{3}{2}} (2s-m)(s-m-\tfrac{1}{2})}{\prod_{m=1}^{s-\frac{3}{2}}(2a-b)(s-m-\tfrac{1}{2})m} = (-1)^{s-\frac{3}{2}}\frac{K}{2}\left(s-\tfrac{1}{2}\right)^2 \frac{(2s+1)!}{\left(\left(s+\tfrac{1}{2}\right)!\right)^2}\left(\frac{|J|}{2K}\right)^{s-\frac{1}{2}} ,
\eeq
where $K = b-2a >0$. The $N$-fold degenerate ground states are
\beq\label{E:gs_specialminus}
{|\psi_0\rangle}_{-} = \begin{cases}
	\tfrac{1}{\sqrt{2}}({|2k-1\rangle}_{-} - {|2k\rangle}_{-})\hspace{30pt} &,\hspace{30pt}s-\tfrac{1}{2}\text{ odd};\\
	\tfrac{1}{\sqrt{2}}({|2k-1\rangle}_{-} + {|2k\rangle}_{-})\hspace{30pt} &,\hspace{30pt}s-\tfrac{1}{2}\text{ even}, 
	\end{cases}
\eeq
for $k=1,2,\dots,N$. The spontaneous breaking of translation invariance is consistent with the Lieb-Schultz-Mattis theorem, which says that the ground state of a translation invariant $s$-spin chain can be gapped \textit{and} translation invariant only if $s-m_0$ is integer, where $m_0$ is the magnetization per site in the ground state \cite{Lieb1961two,mikeska}. For the soliton of length one, the magnetization per site tends to zero in the thermodynamic limit, so $s-m_0$ is noninteger in this limit. 

The calculation for the total spin $\tfrac{1}{2}$ sector is virtually identical, the only difference being that at minimal perturbative order $s-\frac{1}{2}$, the spin $\frac{1}{2}$ sector only has translations of the type
\beq\label{E:trans_plus}
{|2k\rangle}_{+} \longleftrightarrow {|2k+1\rangle}_{+} \hspace{30pt},\hspace{30pt} k=1,2,\dots,N,
\eeq
with ${|N+1\rangle}_{+}={|\overline{1}\rangle}_{+}$ and ${|2N+1\rangle}_{+}={|1\rangle}_{+}$. (Compare with \eqref{E:trans_minus}.) The perturbative mass gap is $2|C^+| = 2|C^-|$, and the $N$-fold degenerate ground states are
\beq\label{E:gs_specialplus}
{|\psi_0\rangle}_{+} = \begin{cases}
	\tfrac{1}{\sqrt{2}}({|2k\rangle}_{+} - {|2k+1\rangle}_{+})\hspace{30pt} &,\hspace{30pt}s-\tfrac{1}{2}\text{ odd};\\
	\tfrac{1}{\sqrt{2}}({|2k\rangle}_{+} + {|2k+1\rangle}_{+})\hspace{30pt} &,\hspace{30pt}s-\tfrac{1}{2}\text{ even}, 
	\end{cases}
\eeq
for $k=1,2,\dots,N$. 

As before, refining our energy denominators to include a first-order $|J|$ term shows that $C$ is of order $|J|$ at the crossover with solitons of length two. (See paragraph following Eqn.\eqref{E:gs_evenspin}.)

\subsection{Solitons of intermediate length}\label{S:any_length}
For solitons of intermediate length $M>1$, translations by one lattice constant occur at relatively high perturbative order. Due to the alternating character of the Néel order, moving the soliton over one site in a \textit{fixed} Néel background cannot result in a translated soliton. (Unless $M=1$, Sec. \ref{S:length_one}.) Solitons of even length, on the one hand, have the following general form
\beq
\cdots\;
\raisebox{-4.5ex}{\includegraphics[]{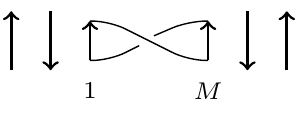}}
\;\cdots
\eeq
where the longest arrows represent maximal spin components of the Néel background. To preserve the right anti-alignment of $s_1$ and $s_M$ with their respective neighbour, translations by one lattice constant require translating the Néel background along, a costly operation. (Unless the soliton's length is larger than the remaining Néel segment, a situation to be discussed in Sec. \ref{S:infinite_length}.) Similarly, solitons of odd length $M>1$ have the general form
\beq
\cdots\;
\raisebox{-4.5ex}{\includegraphics[]{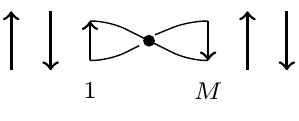}}
\;\cdots
\eeq
where the central dot represents component $0$ ($2s$ even) or $\pm \frac{1}{2}$ ($2s$ odd). Again, the Néel background must be brought along to preserve the right anti-alignment of $s_1$ and $s_M$ with their respective neighbour. For integer spin solitons, of total spin zero, another possibility is to move the soliton by one site in a fixed Néel background, and invert all soliton components. All these transitions occur at relatively high perturbative order.

Obviously, moving the soliton over \textit{two} lattice constants in a fixed Néel background conserves total spin and results in a translated soliton. Such transitions can be performed at comparatively low perturbative order. (See Sec. \ref{S:infinite_length} for comparison.) Because the chain has odd length $N$, the soliton will visit all positions after $N$ steps, and two trips around the chain:
\beq
|\mu\rangle \to |\mu+2\rangle \to \cdots \to |\mu-1\rangle \to |\mu+1\rangle \to \cdots \to |\mu\rangle.
\eeq
At the lowest perturbative order where these translations occur, there are two orthogonal superselection sectors corresponding to the two Néel backgrounds. (If $2s$ is odd and the soliton has odd length, there are four sectors as the central component can be $\pm\frac{1}{2}$.) For each sector, we need to diagonalize a Brillouin-Wigner matrix $w$ analogous to Eqn.\eqref{E:b}. The matrix $w$ is $N\times N$, circulant, of the form
\beq
w=\text{circ}(a_1,a_2,\dots , a_N)=C_s \text{circ}(0,0,1,0,\dots , 0,1,0).
\eeq
Its eigenvalues are
\beq
\epsilon_j = a_1 + a_2 \omega_j + a_3 \omega_j^2 + \dots + a_{N}\omega_j^{N-1},\hspace{40pt} j=0,\dots, N-1,
\eeq
with corresponding eigenvectors $\frac{1}{\sqrt{N}}(1,\omega_j,\omega_j^2,\dots ,\omega_j^{N-1})^{\text{T}}$, where $\omega_j = \exp(i2\pi j /N)$. We find
\beq\label{E:epsilonj}
\epsilon_j = C_s (\omega_j^2 + \omega_j^{N-2})= 2C_s \cos \frac{4\pi j}{N},\hspace{40pt} j=0,\dots, N-1.
\eeq
We will not attempt to compute $C_s$ explicitly, but only whether it is positive or negative for any given $s$, since this is enough to determine the perturbative ground space. The difficulty in calculating $C_s$ explicitly lies in the number of ways in which the transposition operators $S_i^+ S_{i+1}^- + S_i^- S_{i+1}^+$ can be ordered. $C_s$ is a sum over all these different orderings. Each term in that sum is analogous to Eqn.\eqref{E:C_initial}, with a positive amplitude in the numerator, and a product of negative energy denominators. The number of energy denominators is the same in all terms, namely one less than the perturbative order. The sign of $C_s$ is thus determined by the \textit{parity} of the order of perturbation at which the transition occurs
\beq\label{E:C_s_cases}
C_s\begin{cases}
	>0\hspace{20pt}&,\hspace{20pt}\text{odd perturbative order};\\
	<0\hspace{20pt}&,\hspace{20pt}\text{even perturbative order}.
	\end{cases}
\eeq
It is possible that the parity of the minimal perturbative order, and by extension the degenerescence of the ground space, delicately depends on the fine details of the quantum soliton. We assume, however, that on average (including soliton degeneracies) it will be a robust property attached to symmetries which our semiclassical expression is likely to possess as well. 

\subsubsection{Minimal perturbative order}
In order for a soliton to transit over one lattice constant, an amount $\pm s$ of spin from the Néel background must tunnel across it. Let the soliton have arbitrary length $M$, and components $s_1,\dots , s_M$. From Eqn.\eqref{E:adj_trans_res}, the minimal perturbative order at which the component $\pm s$ will transfer is
\beq\label{E:pm_s_transit}
\sum_{i=1}^M |\pm s-s_i|=Ms \mp \sum_{i=1}^M s_i \; .
\eeq
An up-down or down-up pair from the Néel background will therefore transit across a soliton of length $M$ at minimal perturbative order $2Ms$.

\subsubsection{Solitons of even intermediate length}
Solitons of even length have two symmetric halves. As an illustration, consider the soliton of length four, and arbitrary spin $s$:
\[
\cdots\;
\raisebox{-4.5ex}{\includegraphics[]{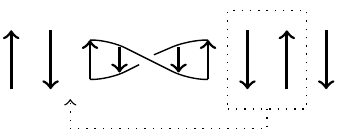}}
\;\cdots
\]
By Eqn.\eqref{E:pm_s_transit}, transferring two adjacent Néel components (boxed) across the soliton requires at least $V^{2Ms}$. Transposing these two maximal components (in order to recover the Néel order to the left of the soliton) requires an additional $V^{2s}$. The minimal perturbative order is thus $2s(M+1)$, where $M+1$ is odd. From Eqn.\eqref{E:C_s_cases}, we get
\beq\label{E:even_length}
(\text{even length})\qquad C_s\begin{cases}
	>0\hspace{20pt}&,\hspace{20pt}2s\text{ odd};\\
	<0\hspace{20pt}&,\hspace{20pt}2s\text{ even}.
	\end{cases}
\eeq
From Eqn.\eqref{E:epsilonj} we conclude that the ground space is fourfold degenerate in each superselection sector if $2s$ is odd, and non-degenerate in each superselection sector if $2s$ is even. In the thermodynamic limit, each sector is gapless with a doubly degenerate ground state for all values of $s$. 

\subsubsection{Solitons of odd intermediate length}\label{S:odd_length}
Solitons of odd length have two antisymmetric halves. As an illustration, consider the soliton of length five:
\[
\cdots\;
\raisebox{-4.5ex}{\includegraphics[]{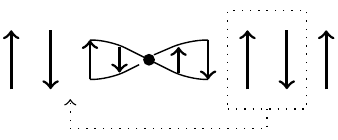}}
\;\cdots
\]
The central dot represents spin component zero if $2s$ is even, and spin component $\pm1/2$ if $2s$ is odd. Notice that once the boxed spins have tunneled across the soliton they already fit in the Néel background, and need not be swapped, in contradistinction to the previous section. The minimal perturbative order for this transition is thus $2Ms$, where $M$ is odd. From Eqn.\eqref{E:C_s_cases}, we get
\beq
(\text{odd length})\qquad C_s\begin{cases}
	>0\hspace{20pt}&,\hspace{20pt}2s\text{ odd};\\
	<0\hspace{20pt}&,\hspace{20pt}2s\text{ even},
	\end{cases}
\eeq
which is identical to the even length case, \eqref{E:even_length}. Again, the ground space is fourfold degenerate in each superselection sector if $2s$ is odd, and non-degenerate in each superselection  sector if $2s$ is even. In the thermodynamic limit, each sector is gapless with a doubly degenerate ground state for all values of $s$.

\subsection{Large solitons}\label{S:infinite_length}
As explained in Section~\ref{S:any_length}, translating a soliton of length $M > 1$ by one lattice constant requires translating the Néel background along. (In some cases, one can flip all soliton components instead. This is a costly operation even for solitons of modest size.) For solitons of length much smaller than the Néel background we argued that it was less expensive to perform translations by two lattice constants in a fixed Néel background. We now consider solitons of length $M$ comparable to the size $N$ of the lattice. For odd values of $N$, these solitons constitute the unperturbed ground states in the region $a\lesssim b$ of parameter space. Since the Néel background is comparatively short, it eventually becomes less expensive to translate it by one lattice constant than translate the soliton a second time. To illustrate our point, let us consider a soliton of large odd length $M$ a fixed fraction of $N$:
\[
\cdots\;
\raisebox{-4.5ex}{\includegraphics[]{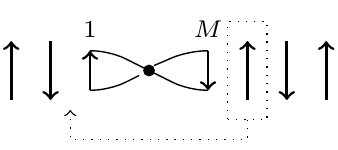}}
\;\cdots
\]
The one site translation $T_1 : |\mu\rangle\to|\mu+1\rangle$ necessitates the concomitant inversion of the Néel background, and can be performed by transiting a Néel component (any one) along the entire chain. From Eqn.\eqref{E:pm_s_transit}, transferring it across the soliton costs 
$Ms + |\sum_{i=1}^M s_i|$, which is equal to $Ms$ for $2s$ even, and $Ms +1/2$ for $2s$ odd. Néel inversion costs an additional 
$\left(\frac{N-M}{2}\right)2s = (N-M)s$. Thus, $T_1$ occurs at minimal perturbative order $\sim Ns$, while $T_2 : |\mu\rangle\to|\mu+2\rangle$ was found to occur at order $2Ms$ in Section~\ref{S:odd_length}. We see that $T_1$ occurs earlier in perturbative theory if $M>N/2$. 
A similar argument can be made when $M$ is  even. We conclude that when $a$ is close enough to $b$, one-site translations occur earlier in perturbation theory than two-site translations. We now briefly summarize the calculation of the perturbative ground state for \textit{large solitons}, i.e. solitons with length $M>N/2$. In all cases, we will find that the ground state is nondegenerate in each superselection sector in the thermodynamic limit. Of course when $2s$ is odd, and since $N$ is odd, there will be \textit{two} superselection sectors mapped onto one another by time reversal, and the ground space will be degenerate in accordance with the Kramers degeneracy theorem. 

\subsubsection{Large even length solitons}
For each Néel background, the $N\times N$ Brillouin-Wigner matrix has the form $w=C\,\text{circ}(0,1,0,\dots , 0,1)$, corresponding to translations $|\mu\rangle\to|\mu\pm 1\rangle$. As before (see Eqn.\eqref{E:epsilon_length_one}) the eigenvalues of $w$ are
\beq\label{E:inf_even_epsilon}
\epsilon_j = C(\omega_j + \omega_j^{N-1})= 2C \cos \frac{2\pi j}{N},\hspace{40pt} j=0,\dots, N-1,
\eeq
with corresponding eigenvectors $\frac{1}{\sqrt{N}}(1,\omega_j,\omega_j^2,\dots ,\omega_j^{N-1})^{\text{T}}$, where $\omega_j = \exp(i2\pi j /N)$. 
The ground state corresponds to $j=0$ when $C<0$, and to $j=\lfloor N/2 \rfloor, \lfloor N/2 \rfloor +1$ when $C>0$. The sign of $C$ is determined by the parity of the minimal perturbative order $\gamma$:
\beq\label{E:inf_even_C}
C=w_{2,1}=\left(\frac{|J|}{2}\right)^{\gamma} \langle 2|S_{1}^- S_{2}^+ \left(\tfrac{Q}{E_1 - H_0}\, S_{1}^- S_{2}^+ \right)^{\gamma-1}|1\rangle
= (-1)^{\gamma -1} \left(\frac{|J|}{2}\right)^{\gamma} \frac{\langle 2|(S_{1}^- S_{2}^+ )^{\gamma} |1\rangle}{\prod_{m=1}^{\gamma-1}|E_1 - e_m|},
\eeq
where $E_1$ and $e_m$ are defined as before, and $E_1 - e_m<0$. The minimal perturbative order $\gamma$ is obtained as above. The one-site translation $T_1$ occurs at the same order as the following transition: the transit of a Néel component across the soliton, plus the complete inversion of the Néel background. The first part occurs at order $Ms + |\sum_{i=1}^M s_i|$. The latter part occurs at order $\left(\frac{N-M-1}{2}\right)2s = (N-M-1)s$. Thus, $T_1$ is of minimal perturbative order $\gamma = (N-1)s+ |\sum_{i=1}^M s_i|$. Since $M$ is even, the soliton's total spin $\sum_{i=1}^M s_i$ is integer, and so is $\gamma$. In the limit $M\to\infty$, our semiclassical expression for the soliton yields  $|\sum_{i=1}^M s_i |\to s$. Numerical results for small to moderate spin values (see Tables~\ref{Spin 1} to~\ref{Spin 7/2}, and Figs.~\ref{F:Spin41half} and~\ref{F:Spin20}) also seem to suggest that for any even $M$, $|\sum_{i=1}^M s_i | = s$ or $|\sum_{i=1}^M s_i | = s -1$ for integer $s$, and $|\sum_{i=1}^M s_i | = s - \frac{1}{2}$ for half-odd $s$. This is the case exactly for the Blume-Capel solitons of the critical line $a=b$, which are of the form $s_i = (-1)^{i} (s-i)$ with $i=1,\dots , 2s-1$. Assuming this to hold as well for the exact quantum soliton of the region $a \lesssim b$, we find 
\beq
\gamma = \begin{cases}
	Ns- \frac{1}{2} \hspace{20pt}&,\hspace{20pt}2s \text{ odd}\\
	Ns, Ns-1	\hspace{20pt}&,\hspace{20pt}2s \text{ even}.
\end{cases}
\eeq
When $2s$ is odd, $\gamma=Ns-\frac{1}{2} = N\left(s-\frac{1}{2}\right) + \frac{N-1}{2}$. Thus, the perturbative ground state is nondegenerate ($C<0$) in each superselection sector if $s-\frac{1}{2}$ and $\frac{N-1}{2}$ are both even or both odd. The perturbative ground state is doubly degenerate in each superselection sector ($C>0$) if exactly one among $s-\frac{1}{2}$ and $\frac{N-1}{2}$ is odd. A new feature of large solitons is that the degeneracy of the perturbative ground state when $2s$ is odd depends on the length $N$ of the entire chain, a global property. When $2s$ is even, our analysis does not allow us to reach a conclusion, unless $a=b$, for then $|\sum_{i=1}^M s_i | = s-1$ if $s$ is odd, and $|\sum_{i=1}^M s_i | = s$ if $s$ is even, implying nondegeneracy ($C<0$) for all integer values of $s$ on the critical line.  

\subsubsection{Large odd length solitons}
For each superselection sector, the $N\times N$ Brillouin-Wigner matrix has the form $w=C\,\text{circ}(0,1,0,\dots , 0,1)$, corresponding to translations $|\mu\rangle\to|\mu\pm 1\rangle$. There are two sectors if $2s$ is even (the two Néel backgrounds), and there are four if $2s$ is odd (two Néel backgrounds, soliton's total spin $\sum_{i=1}^M s_i =\pm\frac{1}{2}$). In each sector, the eigenvalues of $w$ are given by Eqn.\eqref{E:inf_even_epsilon}, and the value of $C$ is given by Eqn.\eqref{E:inf_even_C}. The minimal perturbative order $\gamma$ to reach both one-site translations $T_1 , T_1^{-1}$ was found above to be $Ns+|\sum_{i=1}^M s_i|$. Our semiclassical expression for the soliton of odd length is antisymmetric with respect to the central component. Assuming the exact quantum soliton to possess that symmetry as well, we find
\beq
\gamma = \begin{cases}
	Ns + \frac{1}{2}\hspace{20pt}&,\hspace{20pt}2s\text{ odd};\\
	Ns\hspace{20pt}&,\hspace{20pt}2s\text{ even}.
	\end{cases}
\eeq

When $2s$ is odd, $\gamma=Ns+\frac{1}{2} = N\left(s-\frac{1}{2}\right) + \frac{N+1}{2}$. Thus, the perturbative ground state is nondegenerate ($C<0$) in each superselection sector if $s-\frac{1}{2}$ and $\frac{N+1}{2}$ are both even or both odd. The perturbative ground state is doubly degenerate in each superselection sector ($C>0$) if exactly one among $s-\frac{1}{2}$ and $\frac{N+1}{2}$ is odd. Again, we find that for large solitons, the degeneracy of the perturbative ground state when $2s$ is odd depends on the length $N$ of the entire chain.  When $2s$ is even, $\gamma=Ns$. Thus, the perturbative ground state is nondegenerate ($C<0$) in each superselection sector if $s$ is even, and doubly degenerate in each superselection sector ($C>0$) if $s$ is odd. 

\section{Entanglement spectrum, EE, CE, and correlations}\label{A:EE}
In this appendix, we will compute exactly some quantities related to entanglement in the general perturbative solitonic ground state given in Eqn.\eqref{E:psi_0_res} of the main text. (We do not consider the ground state with broken translation invariance corresponding to the special case of half-odd $s$ and length-one soliton, found in Eqns.\eqref{E:gs_specialminus} and~\eqref{E:gs_specialplus}. All quantities relate to \emph{bipartite} entanglement, that is, they are witness to the entanglement between two non-overlapping intervals on the chain, $A$ and $B$, such that $A\cup B$ is the whole chain.

\subsection{Schmidt decomposition}\label{S:schmidt}
Let $|\psi\rangle$ be a normalized vector in $\mathscr{H}_A \otimes \mathscr{H}_B$. Then there exist orthonormal subsets $\{|u_i\rangle\}\subset \mathscr{H}_A$ and $\{|v_j\rangle\}\subset \mathscr{H}_B$ such that $|\psi\rangle$ may be written as
\beq\label{E:schmidt_appendix}
|\psi\rangle = \sum_{i=1}^Q \sqrt{\lambda_i}|u_i\rangle \otimes |v_i\rangle,
\eeq
where the sum is countable (finite or infinite), $Q\leq \text{min}(\text{dim}\; \mathscr{H}_A,\text{dim}\;\mathscr{H}_B)$, the coefficients $\lambda_i$ are strictly positive, and $\sum_{i=1}^Q \lambda_i=1$. Expression~\eqref{E:schmidt_appendix} is called the \emph{Schmidt decomposition} of $|\psi\rangle$. Note that the coefficients $\lambda_i$ are the eigenvalues of the corresponding reduced density  operators :
\beq
\rho_A = \text{Tr}_B |\psi\rangle\langle\psi| = \sum_{i=1}^Q \lambda_i |u_i\rangle\langle u_i| \qquad , \qquad
\rho_B = \text{Tr}_A |\psi\rangle\langle\psi| = \sum_{i=1}^Q \lambda_i |v_i\rangle\langle v_i| .
\eeq
Finding the Schmidt decomposition of the perturbative solitonic ground state, Eqn.\eqref{E:psi_0_res}, will yield the reduced densities, from which we obtain the entanglement spectrum, EE, and CE.

Let $A$ and $B$ be two intervals such that $\{A,B\}$ is a bipartition of the chain. Define $I$ which we call the seam set, such that  $\mu\in I$ iff the soliton $|\mu\rangle$ has (non Néel) components in \textit{both} $A$ and $B$. Let $A^{\circ}$ stand for the interior of $A$~: $\mu\in A^{\circ}$ iff the soliton part of $|\mu\rangle$ lies entirely within $A$. Define $B^{\circ}$ similarly. Partitioning the sum in Eqn.\eqref{E:psi_0_res} as $\sum_I + \sum_{A^{\circ}} + \sum_{B^{\circ}}$, we obtain the Schmidt decomposition of $|\psi_0\rangle$ with respect to partition $\{A,B\}$,
\beq\label{E:schmidt}
|\psi_0\rangle = \frac{1}{\sqrt{N}}\left( \Big(\sum_{\mu\in I} \omega^{\mu} |\mu\rangle_A |\mu\rangle_B\Big) +  \sqrt{|A^{\circ}|}|A^{\circ}\rangle |\text{Néel}\rangle_B + \sqrt{|B^{\circ}|}|\text{Néel}\rangle_A |B^{\circ}\rangle \right).
\eeq
Let us explain the notation. The state $|\mu\rangle$ in~\eqref{E:psi_0_res} is a tensor product of the states for each spin in the chain. Here $|\cdot\rangle_{A,B}$ naturally stands for the restriction to $A$'s or $B$'s subspace, so that $|\mu\rangle_A$ and $|\text{Néel}\rangle_A$ are vectors that are the tensor product of the spins in the subset $A$ (and similar for $B$). Then, $|A^{\circ}\rangle$ is obtained from $\sum_{\mu=1}^{N}\omega^{\mu}|\mu\rangle_A$ by restricting the sum to $\mu\in A^{\circ}$, and normalizing~: $|A^{\circ}\rangle=|A^{\circ}|^{-1/2}\sum_{\mu\in A^{\circ}}\omega^{\mu}|\mu\rangle_A$. (Here and in what follows, the cardinality of a set $X$ is written $|X|$.) The vector $|B^{\circ}\rangle$ is obtained similarly. Eqn.\eqref{E:schmidt} is formally invariant under the exchange $A\leftrightarrow B$. States corresponding to different classical configurations have vanishingly small overlap, $\langle\eta |\eta'\rangle \sim |J|^{\gamma}$, where as before $|J|\ll 1$ is the coupling constant, and $\gamma$ is the perturbative order at which soliton translations are reached. Thus, the vectors in~\eqref{E:schmidt} effectively constitute an orthonormal Schmidt basis.

The Schmidt rank is defined as the number of terms $Q$ appearing in Eqn.\eqref{E:schmidt_appendix}. Without loss of generality we assume that $A$ is smaller than $B$. Let $R$ be the size of $A$  (so $R < N/2$), and $M$ be the length of the soliton. We must distinguish the cases $M \leq R$ and $M > R$, and the subcases of the latter $M +R\leq N$ and $M + R >N$.

\vspace{10pt}
\noindent\textbf{Case} $M \leq R$. The seam set $I$ in \eqref{E:schmidt} has $2M - 2$ elements, so the Schmidt rank is $2M$. In particular when $M=1$ (Néel defect or minimal soliton), the Schmidt decomposition is seamless ($I=\varnothing$) and the rank is 2. The other extreme subcase is when $M=R$ and $R$ is maximal, $R=(N-1)/2$ : then $I$ has $N-3$ elements, and the Schmidt rank is $N-1$. 

\vspace{10pt}
\noindent\textbf{Case} $M > R$. The set $A^{\circ}$ is empty so $|A^{\circ}\rangle=0$. The set $I$ has $M +R - 1$ element, with a maximum of $N$ :
\beq
| I |=\begin{cases}
	M+R-1 \quad &,\quad M+R \leq N\\
	N \quad &, \quad M +R > N.
	\end{cases}
\eeq
When $I$ has size $N$, the state $|\text{Néel}\rangle_A$ is void as well. The Schmidt rank is $M+R$ (for $M+R \leq N$) with a maximum of $N$ (for $M+R \geq N$). 

\subsection{Reduced density}
The reduced density matrix $\rho_A = \text{Tr}_B | \psi_0\rangle \langle\psi_0|$ is 
\beq\label{E:rho_A}
\rho_A=\frac{1}{N}\left( \sum_{\mu\in I} |\mu\rangle_A {}_A \langle\mu| +|A^{\circ}|\, |A^{\circ}\rangle  \langle A^{\circ}| + |B^{\circ}|\, |\text{Néel}\rangle_A {}_A\langle \text{Néel} |\right) ,
\eeq
with
\beq
\left\{
\begin{aligned}
&|A^{\circ}|=R-M+1 &&, \quad |B^{\circ}|= N-M -R+1 &&\text{ when }M\leq R ; \\
&|A^{\circ}|=0 &&, \quad |B^{\circ}|=N-M -R+1 &&\text{ when }M>R \text{ and } M+R\leq N ; \\
&|A^{\circ}|=0 &&, \quad |B^{\circ}|=0  &&\text{ when }M>R \text{ and } M+R> N.
\end{aligned}
\right.
\eeq
Let us verify proper normalization. In all cases, $\text{Tr}_A \;\rho_A=| I | + |A^{\circ}| + |B^{\circ}|$.

\vspace{10pt}
\noindent\textbf{Case} $M \leq R$. Then $\text{Tr}_A \;\rho_A=\frac{1}{N}\left( (2M -2) +(R-M+1) + (N-M-R+1)\right)=1$.

\vspace{10pt}
\noindent\textbf{Case} $M > R$. If $M+R \leq N$, we have $|A^{\circ}|=0$ and $\text{Tr}_A \;\rho_A=\frac{1}{N}\left( (M +R-1) + (N-M-R+1)\right)=1$. If $M+R > N$, we have $|A^{\circ}|=0, |B^{\circ}|=0$, and $\text{Tr}_A \;\rho_A=\frac{1}{N}\left( N\right)=1$.

\subsection{Entanglement spectrum}
The reduced density \eqref{E:rho_A} may be recast as the partition function at unit temperature ($\beta=1$) of an entanglement spectrum Hamiltonian (or modular Hamiltonian) $H_A$,
\beq\label{E:modular_H}
\rho_A = \sum_n e^{-\xi_n}|n\rangle_A {}_A\langle n | =e^{-H_A}\quad ,
\eeq
whose eigensystem is
\beq
\begin{aligned}\label{E:H_A}
H_A\;|\mu\rangle_A&=\ln N \;|\mu\rangle_A && \qquad(\mu\in I)\\
H_A\;|A^{\circ}\rangle &= \ln \tfrac{N}{R - M +1}\;|A^{\circ}\rangle && \qquad(M\leq R, \text{ void otherwise})\\
H_A\;|\text{Néel}\rangle_A &= \ln \tfrac{N}{N - M -R+1}\;|\text{Néel}\rangle_A && \qquad(M+R \leq N, \text{ void otherwise})
\end{aligned}
\eeq
The ground state of $H_A$ is $|\text{Néel}\rangle_A$, and is gapped from the low-lying state $|A^{\circ}\rangle$.  The higher excited states $|\mu\rangle_A$ form a level with degeneracy $g_{M,R}=| I | = O(M)$,
\beq\label{E:g}
g_{M,R}=\begin{cases}
2M-2 \quad &,\quad M \leq R\\
M+R-1 \quad &,\quad M>R, \;M+R \leq N\\
	N \quad &, \quad M>R, \;M+R > N.
	\end{cases}
\eeq
We will label entanglement energies as follows 
\beq\label{E:xi}
\xi_0 = \ln \tfrac{N}{N - M-R+1}\quad \leq \quad\xi_1 = \ln \tfrac{N}{R-M +1} \quad \leq \quad \xi_2 = \ln N .
\eeq 
A `phase transition of entanglement' of geometrical origin occurs as $R\nearrow N/2$, which closes the gap between $\xi_0$ and $\xi_1$ as the roles of $A$ and $B$ get exchanged. The `energy' gap $|\xi_1 - \xi_0 |$ is a measure of the size difference between $A$ and $B$. Another more physical transition occurs as $M \nearrow R$, at which point the lower excited level $\xi_1$ merges with the upper level $\xi_2$. When $R<M$, all solitons belong to the set $I$, whose size grows with $R$, resulting in entanglement being highly sensitive to $R$. (See the extensive stage of the small-soliton phase, Eqn.\eqref{E:phase1}.) When $R>M$, the size of $I$ plateaus, and so does entanglement entropy. (See the plateaued stage of the small-soliton phase, Eqn.\eqref{E:phase1}.) A third transition is observed as $M+R \nearrow N$, and $\xi_0$ merges with $\xi_2$. Beyond this point, a pure Néel state within $A$ ceases to be a possibility, resulting in all $N$ states within $I$, and saturated entanglement with $B$. (See the saturated stage of the large soliton phase, Eqn.\eqref{E:phase2}.)

\subsection{Entanglement entropy}
The entanglement entropy (EE), or modular thermal average energy (at unit temperature), of subsystem $A$ is defined as
\beq
S_A = -\text{Tr}_A\; \rho_A \ln \rho_A = \sum_n \xi_n e^{-\xi_n} = \langle H_A \rangle_{\beta=1}.
\eeq
Considering multiplicities we have
\beq
S_A=\begin{cases}
\xi_0 e^{-\xi_0}+\xi_1 e^{-\xi_1}+g_{M,R}\,\xi_2 e^{-\xi_2}\quad &,\quad M \leq R\\
\xi_0 e^{-\xi_0}+g_{M,R}\,\xi_2 e^{-\xi_2}\quad &,\quad M>R, \;M+R \leq N\\
g_{M,R}\,\xi_2 e^{-\xi_2}\quad &, \quad M>R, \;M+R > N.
\end{cases}
\eeq
From \eqref{E:g} and \eqref{E:xi} we get
\beq\label{E:EE_discrete}
S_A=\begin{cases}
\frac{N-M -R +1}{N}\ln \frac{N}{N-M -R +1}+\frac{R-M+1}{N}\ln\frac{N}{R-M +1}+\frac{2M -2}{N}\ln N\quad &,\quad \text{region I}\\
\frac{N-M -R +1}{N}\ln \frac{N}{N-M -R +1}+\frac{M +R -1}{N}\ln N\quad &,\quad \text{region II}\\
\ln N\quad &, \quad \text{region III},
\end{cases}
\eeq
with regions I, II, and III as represented in Fig.~\ref{F:regions_res} of the main text. 

\subsection{Correlations}
We now find approximate expressions for the $zz$ correlator over distance $R$,
\beq\label{E:corr_A}
C_{N}^{zz}(R)=\langle\psi_0 | S_i^z S_{i+R}^z |\psi_0\rangle = \frac{1}{N}\sum_{i=1}^N \langle\mu |S_i^z S_{i+R}^z |\mu\rangle,
\eeq
where $|\psi_0\rangle$ is as in Eqn.\eqref{E:psi_0_res}. In the last term, obtained from the translation invariance of $|\psi_0\rangle$, $\mu$ is (any) fixed position on the chain. When separation exceeds the soliton length, $R > M$, we can show, using the symmetries of the semiclassical soliton, that
\beq\label{E:corr_weak_A}
C_{N}^{zz}(R)=(-1)^R s^2 \left(1-\frac{2R}{N}\right).
\eeq
Crucially, as $i$ sweeps over the $N$ sites of the chain, at most one site among $i$ and $i+R$ is on the soliton, the other site being on the Néel background. Because of the complete overturn occuring at the midpoint of the soliton, the sum $\sum_{i,i+R\;\in \text{ soliton}} \langle\mu |S_i^z S_{i+R}^z |\mu\rangle$ vanishes identically (for the semiclassical soliton). For the rest of the sum in~\eqref{E:corr_A}, both $i$ and $i+R$ are on the Néel segment, contributing $(-1)^R s^2$ to the sum if the segment $i,i+1,\dots,i+R$ is Néel, and $(-1)^{R-1} s^2$ if this segment contains the soliton. Thus 
\beq\label{E:corr_terms_A}
C_{N}^{zz}(R)=(-1)^R\; s^2\left( \frac{N-M-R}{N}\right) - (-1)^R\; s^2 \left(\frac{R-M}{N}\right),
\eeq
where the first term comes from Néel segments, and the second term from segments comprising the soliton. From this we obtain~\eqref{E:corr_weak_A}.

We now consider separations inferior to the soliton length, $R<M$, and restrict ourselves to moderate or strong frustration, $M\gg 1$. Using simplifying assumptions, we show that the correlations for this case are
\beq\label{E:corr_largeM_A}
C_{N}^{zz}(R)\sim (-1)^R\; \frac{s^2}{N}\left[N-\frac{2M}{3}-\frac{2R^2}{M}+\frac{2R^3}{3M^2}\right].
\eeq
When both $i$ and $i+R$ are outside the soliton, the segment $i,i+1,\dots,i+R$ can only be Néel, and the contribution to the correlator is identical to the first term of~\eqref{E:corr_terms_A}. For the rest of the calculation, we approximate the soliton profile with a linearized one,
\beq
s_j \approx (-1)^j \frac{2s}{M} \left(\frac{M}{2}-j\right)\qquad , \qquad j=1,\dots,M-1.
\eeq
As observed in Section~\ref{S:free_theory_res}, very large (semiclassical) solitons, $M\lesssim N$, have nearly this form. (See also Figs.\ref{F:Spin20}.\textbf{f} and~\ref{F:Spin41half}.\textbf{f}.) When exactly one site among $i,i+R$ is on the soliton (denoted $i\in I_1$ below), the contribution to the correlator is 
\beq\label{E:one_site_on_soliton}
\frac{1}{N}\sum_{i\in I_1} \langle\mu |S_i^z S_{i+R}^z |\mu\rangle\approx (-1)^R \; \frac{2s^2}{N}\left(\frac{2}{M}\right)\sum_{j=1}^R \left(\frac{M}{2}-j\right) .
\eeq
When sites $i,i+R$ are on the soliton simultaneously (denoted $i\in I_2$), we find
\beq\label{E:sites_on_soliton}
\frac{1}{N}\sum_{i\in I_2} \langle\mu |S_i^z S_{i+R}^z |\mu\rangle\approx(-1)^R\; \frac{2s^2}{N}\left(\frac{2}{M}\right)^{\hspace{-3pt}2} \hspace{5pt}\sum_{j=1}^{M-R}\left(\frac{M}{2}-j\right)\left(\frac{M}{2}-j-R\right).
\eeq 
Combining~\eqref{E:one_site_on_soliton},~\eqref{E:sites_on_soliton} and the first term of~\eqref{E:corr_terms_A}, we obtain~\eqref{E:corr_largeM_A} in the limit $M\gg 1$.

\end{document}